\documentclass[11pt,letterpaper]{article}
\usepackage{jheppub}
\usepackage{graphicx, epsfig} %include figure files
\usepackage{amsmath,amssymb,amsfonts,dsfont,mathrsfs,amsthm,mathtools}
\usepackage{slashed}
\usepackage{bm} %include bold math: \bm{} creates bold letters in math mode
\usepackage{color}
\usepackage[usenames]{xcolor}
\usepackage{hyperref}
\usepackage{siunitx}
\hypersetup{colorlinks=true,urlcolor=blue,linkcolor=magenta,citecolor=blue,filecolor=blue}
\usepackage[normalem]{ulem}
\usepackage{array}
\usepackage{booktabs}

\newcommand{\diff}[1]{\text{d}#1}

\newcommand{\Lie}{\mathcal{L}}

\begin{document}

%%%%%%%
\title{Gravitational instantons with conformally coupled scalar fields}
%%%%%%%

% \author{Jos\'e Barrientos}
% \email{barrientos@math.cas.cz}
% \affiliation{Institute of Mathematics of the Czech Academy of Sciences, \v{Z}itn\'a 25, 11567 Praha 1, Czech Republic}
% \affiliation{Departamento de Ense\~nanza de las Ciencias B\'asicas,
% Universidad Cat\'olica del Norte, Larrondo 1281, Coquimbo, Chile}

% \author{Adolfo Cisterna}
% \email{adolfo.cisterna.r@mail.pucv.cl}
% \affiliation{Sede Esmeralda, Universidad de Tarapacá, Av. Luis Emilio Recabarren 2477, Iquique, Chile}

% \author{Crist\'obal Corral}
% \email{crcorral@unap.cl}
% \affiliation{Instituto de Ciencias Exactas y Naturales, Facultad de Ciencias, Universidad Arturo Prat, Avenida Arturo Prat Chac\'on 2120, 1110939, Iquique, Chile}

% \author{Marcelo Oyarzo}
% \email{moyarzo2016@udec.cl}
% \affiliation{Departamento de F\'isica, Universidad de Concepci\'on, Casilla, 160-C, Concepci\'on, Chile}

\author[a,b]{Jos\'e Barrientos,}
\author[c]{Adolfo Cisterna,} 
\author[d]{Crist\'obal Corral,}
\author[e]{and Marcelo Oyarzo}

\emailAdd{barrientos@math.cas.cz}
\emailAdd{adolfo.cisterna.r@mail.pucv.cl}
\emailAdd{crcorral@unap.cl}
\emailAdd{moyarzo2016@udec.cl}
\affiliation[a]{Institute of Mathematics of the Czech Academy of Sciences, \v{Z}itn\'a 25, 11567 Praha 1, Czech Republic}
\affiliation[b]{Departamento de Ense\~nanza de las Ciencias B\'asicas,
Universidad Cat\'olica del Norte, Larrondo 1281, Coquimbo, Chile}
\affiliation[c]{Sede Esmeralda, Universidad de Tarapacá, Av. Luis Emilio Recabarren 2477, Iquique, Chile}
\affiliation[d]{Instituto de Ciencias Exactas y Naturales, Facultad de Ciencias, Universidad Arturo Prat, Avenida Arturo Prat Chac\'on 2120, 1110939, Iquique, Chile}
\affiliation[e]{Departamento de F\'isica, Universidad de Concepci\'on, Casilla, 160-C, Concepci\'on, Chile}

\abstract{
%\begin{abstract}
We present novel regular Euclidean solutions to General Relativity in presence of Maxwell and conformally coupled scalar fields. In particular, we consider metrics of the Eguchi-Hanson and Taub-NUT families to solve the field equations analytically. The solutions have nontrivial topology labeled by the Hirzebruch signature and Euler characteristic that we compute explicitly. We find that, although the solutions are locally inequivalent with the original (anti-)self-dual Eguchi-Hanson metric, they have the same global properties in the flat limit. We revisit the Taub-NUT solution previously found in the literature, analyze their nuts and bolts structure, and obtain the renormalized Euclidean on-shell action as well as their topological invariants. Additionally, we discuss how the solutions get modified in presence of higher-curvature corrections that respect conformal invariance. In the conformally invariant case, we obtain novel Eguchi-Hanson and Taub-NUT solutions and demonstrate that both Euclidean on-shell action and Noether-Wald charges are finite without any reference to intrinsic boundary counterterms.
%\end{abstract}
}
%%%%%%

\maketitle

\section{Introduction}

%Instantons are remarkable vacuum solutions of Euclidean Yang-Mills equations, labeled by a topological number: the Chern-Pontryagin index. They exhibit nonperturbative effects in the path integral formulation appearing as the leading quantum correction to classical systems and they can be used to compute tunneling amplitudes between topologically inequivalent vacua (for reviews see~\cite{Coleman:1978ae,Gross:1980br,Schafer:1996wv}). Their existence is related to the fermionic axial anomaly and nonpertubative mass generation of hadrons, portraying the nontrivial vacuum structure of gauge theories~\cite{tHooft:1976rip}. When instantons are (anti-)self dual, the Yang-Mills action become proportional to the Chern-Pontryagin index, saturating the BPS bound~\cite{Belavin:1975fg}. Thus, since their discovery, they have become objects of huge interest in theoretical and mathematical physics due to their wide range of applications.  

Gravitational instantons are remarkable solutions to Euclidean gravity that have attracted a lot of interest since the seminal works of Hawking et al.~\cite{Hawking:1976jb,Gibbons:1978tef,Hawking:1978ghb} and Eguchi-Hanson~\cite{Eguchi:1978xp,Eguchi:1978gw}. These geometries are characterized by their nontrivial topology measured by the Hirzebruch signature and Euler characteristic; both being topological invariants related to the difference between harmonic self dual and anti-self dual forms in the middle dimension and the number of genus/boundaries of the manifold, respectively (for a review see~\cite{Eguchi:1980jx}). In some cases, they possess either (anti-)self dual Riemann or Weyl tensor, and it is expected that they will play a crucial role in the path integral approach to quantum gravity, similar to their Yang-Mills counterparts in quantum field theory. Their existence is related to the axial anomaly in curved spacetimes~\cite{Alvarez-Gaume:1983ihn,Witten:1985xe} and they have motivated different applications from theoretical physics to differential geometry~\cite{Linshaw:2017bpf,Hashemi:2018jbv,Li2019}.

An interesting example of gravitational instantons can be obtained from the analytic continuation of the Taub-Newman-Tamburino-Unti metric~\cite{Taub:1950ez,Newman:1963yy}---hereafter referred to as Taub-NUT. Depending on whether its set of fixed points is zero or two-dimensional~\cite{Page:1978hdy}, the solution is said to have a nut or a bolt, respectively~\cite{Gibbons:1979xm}. The Taub-NUT metric was originally found as a stationary extension of the Schwarzschild black hole~\cite{Taub:1950ez,Newman:1963yy} and it is continuously connected to the latter in the limit when the new parameter---the NUT charge---vanishes. This solution is usually interpreted as a gravitational dyon since the NUT charge plays the role of a gravitomagnetic monopole, sourcing the magnetic part of the Weyl tensor~\cite{Lynden-Bell:1996dpw,Bicak:2000ea,Miskovic:2009bm,Araneda:2016iiy}. In presence of Maxwell fields, the first charged Taub-NUT solution was obtained by Brill in~\cite{PhysRev.133.B845} and different aspects of their thermodynamics have been recently studied~\cite{Mann:2020wad,Abbasvandi:2021nyv}.\footnote{In presence of nonlinear conformal electrodynamics~\cite{Bandos:2020jsw}, Taub-NUT solutions have been found in~\cite{BallonBordo:2020jtw,Flores-Alfonso:2020nnd,Zhang:2021qga}.} Holographic aspects of rotating Taub-NUT-AdS spaces have been studied through the Kerr/CFT correspondence~\cite{Sakti:2019udk,Sakti:2019zix,Sakti:2019krw,Sakti:2020jpo} and it has been demonstrated that this solution describes holographic fluids with nontrivial vorticity~\cite{Leigh:2011au,Leigh:2012jv,Caldarelli:2012cm,Mukhopadhyay:2013gja,Kalamakis:2020aaj}. 

It is well-known that the Taub-NUT space is endowed with a string-like singularity dubbed the Misner string~\cite{Misner:1963fr}. The latter represents the gravitational analog of the Dirac string and its position can be made unobservable through a proper choice of coordinates. This condition, however, implies that the time coordinate must be periodic and it endows the spacetime with close time-like curves in Lorentzian signature~\cite{Misner:1963fr}. Due to this, the Taub-NUT metric is usually studied in Euclidean signature by virtue of their close resemblance with instantons in Yang-Mills theory. Nevertheless, it is worth mentioning that recent developments on the Lorentzian Taub-NUT spacetime has attracted a lot of interest~\cite{Hennigar:2019ive,Chen:2019uhp,Kalamakis:2020aaj,BallonBordo:2020mcs,Cano:2021qzp,Abbasvandi:2021nyv}, since it has been shown that the Misner string is fully transparent to geodesic observers~\cite{Clement:2015cxa,Clement:2015aka}.  

As Hawking and Hunter shown in~\cite{Hawking:1998jf}, the origin of gravitational entropy can be traced to obstructions to foliation of topologically nontrivial Euclidean spaces with a scalar function that provides the unitary Hamiltonian evolution. For black holes, these obstructions correspond to their event horizon. In the Taub-NUT case, in turn, the Misner string represents an additional source of entropy~\cite{Hawking:1998jf,Garfinkle:2000ms}, breaking the standard $S=\frac{A}{4G}$ law of black hole thermodynamics~\cite{Astefanesei:2004ji}. Indeed, in anti de Sitter (AdS) space, the contribution of the Misner string becomes divergent and a renormalization scheme is needed (see for instance~\cite{Hawking:1998ct,Emparan:1999pm,Mann:1999pc,Ciambelli:2020qny}). In all cases, the renormalized entropy yields a consistent first law of thermodynamics, even by considering their extended phase space~\cite{Johnson:2014xza,Johnson:2014pwa}.

Another interesting example of gravitational instantons found in general relativity is the asymptotically flat (anti-)self dual space obtained by Eguchi and Hanson in Refs.~\cite{Eguchi:1978xp,Eguchi:1978gw}. This solution bears close resemblance with the pseudoparticle solution to Euclidean Yang-Mills equations obtained by Belavin-Polyakov-Schwartz-Tyupkin in~\cite{Belavin:1975fg}. In absence of the cosmological constant, this space has (anti-)self dual Riemann tensor and, by virtue of Bianchi identities, it solves the Einstein's equations automatically. These solutions have vanishing Euclidean on-shell action and nontrivial topological invariants and some of their extensions have been studied in~\cite{Xiao_2004,Chen:2020org,Corral:2021xsu}. Thus, it is also expected that they will play an important role in quantum gravity. 

In theories beyond general relativity, gravitational instantons have been focus of main interest as well. In supergravity, for instance, it was shown that gravitational instantons break the global $U(1)$ axial symmetry of spin-$3/2$ zero modes, giving rise to helicity-changing amplitudes~\cite{Hawking:1978ghb}. In the low energy limit of string theory, different Taub-NUT solutions have been obtained in~\cite{Burgess:1994kq,Johnson:1994ek,Johnson:1994nj}. Higher-curvature corrections to general relativity are also endowed with this class of solutions~\cite{Strominger:1984zy,Dehghani:2005zm,Dehghani:2006aa,Hendi:2008wq,Bueno:2018uoy,Corral:2019leh,Corral:2021xsu} and, when invariance under Weyl rescaling is present, it has been shown that Euclidean on-shell action of conformal gravity and conserved charges thereof in AdS are finite without any reference of boundary counterterms~\cite{Corral:2021xsu} (see Ref.~\cite{Grumiller:2013mxa} for details).

Since the seminal works of Israel, Carter and Wald \cite{Carter:1971zc,Israel:1967wq,Israel:1967za,Wald:1971iw}, showing the Kerr-Newman solution to be the unique rotating axially-symmetric black hole of Einstein-Maxwell theory, the study of black holes with other possible characteristics beyond the mass $M$, electromagnetic charges $Q_{e,m}$ and angular momentum $J$ has been a cornerstone in the study of classical gravity. The Kerr solution successfully provides a relativistic model for the gravitational field of a rotating central mass possessing angular velocity. Thus, its uniqueness has lead to the conclusion that a rotating collapsing star losses all possible traces of its individual features, settling down to an equilibrium state uniquely characterized by the aforementioned set of parameters. This statement is materialized by Wheeler's conjecture ``black hole have no hair'': For any type of energy-matter distribution, the outcome of its gravitational collapse will be a Kerr black hole exclusively described by its mass, charge and angular momentum. Any other quantities---for which the adjective of hair represents a metaphor---are either eaten up or expelled out during the collapse and do not take part in the final description of the black hole. 
However, the no-hair conjecture is intrinsically model dependent and the search for scalar hair has been historically investigated, probing itself as a fertile ground for the construction of new black hole solutions with possible astrophysical relevance.

Scalar-tensor theories have provided a rich reservoir of hairy black hole solutions, systematically circumventing no-hair theorems by means of different approaches~\cite{Herdeiro:2015waa}. The archetypical model is given by Einstein gravity supplemented by conformally coupled scalar fields for which several black hole solutions, beyond spherical symmetry, have been found~\cite{Bekenstein:1974sf,Bocharova:1970skc,Martinez:2002ru,Martinez:2005di,Charmousis:2009cm,Astorino:2013xxa,Bardoux:2013swa,Astorino:2014mda,Anabalon:2009qt,Cisterna:2021xxq,Astorino:2013xc,Astorino:2013sfa,Caceres:2020myr,Anabalon:2012tu,Ayon-Beato:2015ada,Barrientos:2022avi}. 
Even in the case of gravitational instantons, where only few numerical gravitational instantons with nontrivial scalar field are known~\cite{Brihaye:2016lsx,Arratia:2020hoy}, this model proved itself fruitful providing a remarkable exception of an analytic gravitational instanton~\cite{Bardoux:2013swa,Bhattacharya:2013hvm}. In those references, the authors found analytical Taub-NUT solutions in Lorentzian and Euclidean signature with a nontrivial scalar field, with and without electric charge.

The aim of this work is to extend the space of solutions of general relativity with a conformally coupled scalar field in the context of exact gravitational instantons.\footnote{In the Minkowski background, Yang-Mills instantons in this theory were found in Ref.~\cite{Eguchi:1976db} and, for AdS, scalar instantons were obtained in~\cite{deHaro:2006ymc}.} To this end, we find a generalization of the original Eguchi-Hanson metric with a nontrivial scalar profile. We obtain $U(1)$ gauge fields as solutions to the field equations possessing divergent pieces at the action level. Nevertheless, the introduction of topological terms with fixed coupling constant renders the Maxwell action finite while setting (anti-)self-dual configurations as the ground state of the theory. In the conformally invariant case, namely, when the Einstein sector decouples from the scalar-tensor one, the theory becomes a particular case of Brans-Dicke gravity enjoying conformal symmetry. Moreover, the Euclidean on-shell action, alongside the Noether-Wald charges, are finite without any reference to intrinsic boundary counterterms; similar to the metric formulation of conformal gravity~\cite{Grumiller:2013mxa,Anastasiou:2020mik,Anastasiou:2021tlv,Corral:2021xsu}. We study topological properties of previously found solutions of the Taub-NUT family~\cite{Bardoux:2013swa} and we show that higher-curvature corrections allows one to construct solutions of the Eguchi-Hanson type possessing a negative cosmological constant and a positive curvature K\"ahler base manifold, something forbidden in their absence.

The manuscript is organized as follows: In Sec.~\ref{sec:thetheory}, we present the theory under consideration, alongside the field equations. In Sec.~\ref{sec:solutions}, we solve the field equations and obtain the charged generalized Eguchi-Hanson metric and revisit the Taub-NUT solution presented in Ref.~\cite{Bardoux:2013swa}. In Sec.~\ref{sec:higher-curvature}, we obtain solutions in presence of higher-curvature terms that respect invariance under Weyl rescalings. Section~\ref{sec:conformal} is devoted to present novel gravitational instantons in a particular class of Brans-Dicke gravity possessing conformal invariance and we show that the Noether-Wald charges are finite without intrinsic boundary counterterms. Finally, in Sec.~\ref{sec:conclusions} we provide concluding remarks and possible future directions.

%Some extensions of the Eguchi-Hanson metric have been explored in~\cite{Xiao_2004,Chen:2020org} and it has been recen

\section{Conformally coupled scalar field\label{sec:thetheory}}

The gravitational dynamics we study throughout this manuscript is dictated by the Einstein-Maxwell theory augmented by a conformally coupled scalar field. We will work mainly in Euclidean signature since we are interested in instantons. Thus, we define the Euclidean action with a minus sign in front such that, to first order in the saddle point approximation, it gives $\ln Z \approx -I_E$, where $Z$ is the partition function. The action principle is then given by
\begin{align}\label{Ibulk}
    I_{\rm bulk} &= \int_{\mathcal{M}}\diff{^4x}\sqrt{|g|}\left[\kappa\left(R-2\Lambda \right) - \frac{1}{2}g^{\mu\nu}\nabla_\mu\phi\nabla_\nu\phi - \frac{1}{12}R\phi^2 - \alpha\phi^4 -\frac{1}{4}F^2 - \Theta\tilde{F}F  \right]\,,
    %\label{IGHY}
    %I_{\rm GHY} &= 2\int_{\partial\mathcal{M}}\diff{^3x}\sqrt{|h|}\left[\kappa-\frac{\phi^2}{12}\right]K\,,
    %\label{Ict}
    %I_{\rm ct} &= \int_{\partial\mathcal{M}}\diff{^3x}\sqrt{|h|}\left(\zeta_1 + \zeta_2\mathcal{R} + \zeta_3\mathcal{R}^2 + \ldots \right)\,.
\end{align}
where $g=\det g_{\mu\nu}$, $\kappa=\left(16\pi G \right)^{-1}$ is the gravitational constant and $\alpha$ is a dimensionless parameter controlling the quartic potential for the scalar field. In the Maxwell sector, we have defined the invariants $F^2 \equiv F_{\mu\nu}F^{\mu\nu}$ and $\tilde{F}F\equiv\tilde{F}_{\mu\nu}F^{\mu\nu}=\tfrac{1}{2}\varepsilon_{\mu\nu\lambda\rho}F^{\mu\nu}F^{\lambda\rho}$, where $\varepsilon_{\mu\nu\lambda\rho}$ is the Levi-Civita tensor. The last term in Eq.~\eqref{Ibulk} is proportional to the Chern-Pontryagin index of $U(1)$. In Yang-Mills theory, the latter represents the contribution of the $\theta$-vacuum to physical observables at the quantum level, such as the electric dipole moment of the neutron (for a review see~\cite{Peccei:2006as}). In this case, the Pontryagin density plays a two-fold role: (i) setting (anti-)self-dual Maxwell fields as the ground state of the theory~\cite{Miskovic:2009bm,Araneda:2016iiy}, and (ii) renormalizing the Euclidean on-shell action for Maxwell fields over the Eguchi-Hanson space.

The field equations can be obtained by performing stationary variations with respect to the metric, Maxwell and scalar field, giving
\begin{subequations}\label{eom}
\begin{align}\label{eomg}
\mathcal{E}_{\mu\nu} &\equiv 2\kappa\left(G_{\mu\nu} + \Lambda g_{\mu\nu}\right) - T_{\mu\nu} = 0,\\
\label{eoma}
\mathcal{E}^\mu &\equiv \nabla_\nu F^{\nu\mu} = 0\,,\\
\label{eomp}
\mathcal{E} &\equiv \Box\phi - \frac{1}{6}R\phi - 4\alpha\phi^3=0\,,
\end{align}
\end{subequations}
respectively, where $G_{\mu\nu}=R_{\mu\nu} - \tfrac{1}{2}g_{\mu\nu}R$ is the Einstein tensor and $T_{\mu\nu}=T_{\mu\nu}^{(\phi)}+T_{\mu\nu}^{(A)}$ is the total stress-energy tensor whose scalar and Maxwell pieces are respectively defined as
\begin{align}\label{Tphimunu}
    T_{\mu\nu}^{(\phi)} &= \nabla_\mu\phi\nabla_\nu\phi - \frac{1}{2}g_{\mu\nu}\nabla_\lambda\phi\nabla^\lambda\phi + \frac{1}{6}\left[g_{\mu\nu}\Box - \nabla_\mu\nabla_\nu + G_{\mu\nu} \right]\phi^2 - \alpha g_{\mu\nu}\phi^4\,, \\
    \label{TAmunu}
    T_{\mu\nu}^{(A)} &= F_{\mu\lambda}F_{\nu}{}^{\lambda} - \frac{1}{4}g_{\mu\nu}F_{\lambda\rho}F^{\lambda\rho}\,.
\end{align}
The tensor-scalar and Maxwell sector of the theory are invariants up-to-boundary terms under Weyl rescaling, namely, $g_{\mu\nu}\to\Omega^2(x)g_{\mu\nu}$, $\phi\to\Omega^{-1}(x)\phi$, and $F_{\mu\nu}\to F_{\mu\nu}$. This, in turn, implies that the trace of the energy-momentum tensor is proportional to the Klein--Gordon equation and it vanishes on-shell. Taking the trace on Eq.~\eqref{eomg}, one concludes that the Ricci scalar is $R=4\Lambda$\,. 

In the following, we extend the space of solutions of the theory~\eqref{Ibulk} by considering gravitational analog of instantons in Riemannian signature and studying some of their properties.

\section{Gravitational instantons\label{sec:solutions}}

In this section, we solve analytically the field equations~\eqref{eom} to obtain solutions of the Eguchi-Hanson family with a nontrivial scalar and Maxwell fields. We also revisit the Euclidean Taub-NUT solution found in Refs.~\cite{Bardoux:2013swa,Bhattacharya:2013hvm} to study their global properties. The renormalized Euclidean on-shell action of the latter is obtained by introducing intrinsic boundary counterterms that render its value finite in spaces with locally Euclidean AdS asymptotics. This prescription, however, is ambiguous for the Eguchi-Hanson solution. Nevertheless, we show that there is a remarkable exception in the conformally invariant case, i.e., in absence of the Einstein and cosmological terms, that leads to a particular Brans-Dicke theory with finite conserved charges and Euclidean on-shell action without any reference to intrinsic boundary counterterms. 

\subsection{Eguchi-Hanson}

To solve the field equations, we consider a metric ansatz based on the Eguchi-Hanson metric with a $U(1)$ fibration of K\"ahler manifold with constant curvature $\gamma=\pm1,0$. In particular, we focus on the line element
\begin{align}\label{EHansatz}
    \diff{s^2} = \frac{r^2f(r)}{4}\left(\diff{\tau} + B_{(\gamma)} \right)^2 + \frac{\diff{r^2}}{f(r)} + \frac{r^2}{4}\diff{\Sigma^2_{(\gamma)}}\,,
\end{align}
where the K\"ahler potential 1-form is given by
\begin{align}\label{basemanifold}
    B_{(\gamma)} = \left\{ \begin{matrix} \cos\vartheta\diff{\varphi} & \mbox{when} & \gamma=1 & \mbox{and} &  \diff{\Sigma^2_{(\gamma=1)}} = \diff{\vartheta}^2 + \sin^2\vartheta\diff{\varphi^2}\,, \\
    \frac{1}{2}\left(\vartheta\diff{\varphi}-\varphi\diff{\vartheta}\right) & \mbox{when} & \gamma=0 & \mbox{and} & \hspace*{-1cm} \diff{\Sigma^2_{(\gamma=0)}} = \diff{\vartheta}^2 + \diff{\varphi^2}\,, \\
    \cosh\vartheta\diff{\varphi} & \mbox{when} & \gamma=-1 & \mbox{and} &  \diff{\Sigma^2_{(\gamma=-1)}} = \diff{\vartheta}^2 + \sinh^2\vartheta\diff{\varphi^2}\,.
    \end{matrix}     \right.
\end{align}
Since the metric $\diff{\Sigma_{(\gamma)}^2}$ is K\"ahler, its associated real fundamental $2$-form, say $\Omega_{(\gamma)}$, is closed. Thus, the Poincaré lemma implies that locally it can be written in terms of the K\"ahler potential as $\Omega_{(\gamma)}=\diff{B_{(\gamma)}}$. For the scalar and Maxwell fields we assume the same isometry group underlying the metric~\eqref{EHansatz}.  

The field equations~\eqref{eom} admit the following metric function, scalar and Maxwell field as solutions, that is
\begin{align}\label{sol1}
    f(r) &= \gamma - \frac{\Lambda r^2}{6} + \frac{b}{r^2}\,, &  \phi(r) &= \frac{1}{r^2} \sqrt{\frac{b}{\alpha}}\,, & A = \left[\frac{q}{r^2}-\frac{b(72\,\alpha\,\kappa+\Lambda)}{576\,q\,\alpha}\;r^2\right]\left(\diff{\tau} + B_{(\gamma)} \right)\,,
\end{align}
where $b$ and $q$ are integration constants, $A=A_\mu\diff{x^\mu}$ is the $1$-form Maxwell field, and $q,\alpha\neq0$. Let us focus on the case with $\gamma=1$ first. For this choice, the metric~\eqref{EHansatz} is locally invariant under the action of the symmetry algebra $\mathfrak{su}(2)\otimes \mathfrak{u}(1)$ and it can be written in terms of the left-invariant Maurer-Cartan forms of $SU(2)$, $\sigma_i$, through
\begin{align}\label{Eguchi-Hanson-ansatz}
\diff{s^2} = \frac{\diff{r^2}}{f(r)} + r^2 \left( \sigma_1^2 + \sigma^2_2 + {f(r)} \sigma_3^2\right)\,.
\end{align}
These one-forms satisfy $\diff{\sigma_i}=\epsilon_{ijk}\sigma_j\wedge\sigma_k$ and, using the basis of Euler angles with $0\leq\theta\leq\pi$, $0\leq\phi\leq 2\pi$ and $0\leq\tau\leq 4\pi$, they can be explicitly represented as
\begin{subequations}\label{MCformsu2}
\begin{align}
\sigma_1 &= \frac{1}{2}\left(\sin\tau \diff{\vartheta} - \sin\vartheta  \cos\tau  \diff{\varphi}\right)\,, \\
\sigma_2 &= -\frac{1}{2}\left(\cos\tau \diff{\vartheta} + \sin\vartheta  \sin\tau \, \diff{\varphi}\right)\,, \\
\sigma_3 &= \frac{1}{2}\left(\diff{\tau} +\cos \vartheta \diff{\varphi}\right)\, . 
\end{align}
\end{subequations}
The existence of a bolt at $r=r_b>0$ defined through $f(r_b)=0$ implies that: (i) $\alpha,\Lambda,b<0$ or (ii) $\alpha,\Lambda,b>0$. When the cosmological constant is negative, we define $\ell^{-2}=-\Lambda/6$ and $Q=2q$ and the condition $r_b^2+b<0$ must hold. Then, the solutions in Eq.~\eqref{sol1} become
\begin{align}\label{sol1a}
    f(r) &= 1 + \frac{r^2}{\ell^2} + \frac{b}{r^2}\,, & \phi(r) &= \frac{1}{r^2} \sqrt{\frac{b}{\alpha}}\,, & A = \left[\frac{Q}{r^2} - \frac{b\left(3\alpha - 4\pi G\ell^{-2} \right)}{96\pi G \alpha Q}  \;r^2\right]\sigma_3\,,
\end{align}
where $\kappa=(16\pi G)^{-1}$ has been used. This solution has a conical singularity at $r=r_b$ for the standard period of the Euclidean time defined through Euler's angles. This can be eliminated by demanding their right periodicity to be
\begin{align}\label{betasolEH}
    \beta_\tau = -\frac{4\pi r_b^2}{r_b^2+2b}\,,
\end{align}
where, recall, $r_b^2+2b<0$, such that the period of the Euclidean time is positive. 

The Weyl tensor associated to this solution is globally (anti-)self dual if and only if $\ell\to\infty$, as it can be seen from the nonvanishing invariant
\begin{align}
    \left(W^{\mu\nu}_{\lambda\rho}\pm\tilde{W}^{\mu\nu}_{\lambda\rho}\right)\left(W_{\mu\nu}^{\lambda\rho}\pm\tilde{W}_{\mu\nu}^{\lambda\rho}\right) = \frac{384}{\ell^4}\,.
\end{align}
Here, $W^{\mu\nu}_{\lambda\rho}$ is the Weyl tensor and $\tilde{W}_{\mu\nu\lambda\rho}=1/2\,\varepsilon_{\mu\nu\sigma\tau}W^{\sigma\tau}_{\lambda\rho}$ its dual. Additionally, it is worth mentioning that this solution is not asymptotically locally Euclidean AdS, as the type I Eguchi-Hanson metric studied in~\cite{Chen:2020org}. This can be seen by checking that the curvature invariant constructed out of the Weyl tensor  
\begin{align}
    W^{\mu\nu}_{\lambda\rho}W^{\lambda\rho}_{\mu\nu} = \frac{96}{\ell^4} + \frac{96\,b^2}{r^8}\,,
\end{align}
is nonvanishing asymptotically. This is different from the asymptotic behavior of the Euclidean Taub-NUT/Bolt-AdS which is indeed asymptotically locally AdS. 

The global properties of the solution are labeled by the Hirzebruch signature $\tau(\mathcal{M})$ and Euler characteristic $\chi(\mathcal{M})$. For this solution, these topological invariants are respectively given by
\begin{align}\notag 
   \tau(\mathcal{M})  &= \frac{1}{96\pi^2} \int_{\mathcal{M}} d^4x\,\sqrt{g}\, \varepsilon_{\mu\nu \lambda\rho }R_{\sigma\tau}^{\lambda\rho}R^{\sigma\tau\mu\nu } \\
    \label{signature1}
   &\quad - \frac{1}{4\pi^2}\int_{\partial\mathcal{M}}   \diff{^3x}\sqrt{h}\;n_\mu\varepsilon^{\mu\nu\lambda\rho}K^\sigma_\nu\nabla_\lambda K_{\rho\sigma}  = -1 - \frac{2r_b^2}{\ell^2}\,,\\
    \notag
 \chi(\mathcal{M}) &=\frac{1}{32\pi^2}\int_{\mathcal{M}} d^4x\,\sqrt{g}\left( R_{\lambda\rho}^{ \mu\nu}R^{\lambda\rho}_{\mu\nu}-4R^\mu_\nu R^\nu_\mu +R^2\right) \\
 &\quad+ \frac{1}{8\pi^2}\int_{\partial\mathcal{M}}\diff{^3x}\;\sqrt{h}\;\delta^{\alpha\beta\gamma}_{\mu\nu\lambda}K^\mu_\alpha\left(\frac{1}{2}\mathcal{R}^{\nu\lambda}_{\beta\gamma} - \frac{1}{3}K^\nu_\beta K^\lambda_\gamma \right) = 2\,, \label{eulerch1}
\end{align} 
In the flat limit, namely $\ell\to\infty$, these topological invariants coincide with the ones of the asymptotically Euclidean Eguchi-Hanson space; they are $\tau(\mathcal{M})=-1$ and $\chi(\mathcal{M})=2$; although the solutions are manifestly locally inequivalent.

The second term of the Maxwell field in Eq.~\eqref{sol1a} produces a divergence in the electromagnetic fields as $r\to\infty$, since the $2$-form field strength $F=\diff{A}$ is
\begin{align}
    F = -\left[\frac{2Q}{r^3} + \frac{b\left(3\alpha - 4\pi G\ell^{-2} \right)}{48\pi G \alpha Q}  \;r\right]\diff{r}\wedge\sigma_3 + \left[\frac{2Q}{r^2} - \frac{b\left(3\alpha - 4\pi G\ell^{-2} \right)}{48\pi G \alpha Q}  \;r^2\right]\sigma_1\wedge\sigma_2\,.
\end{align}
The divergent term does not allow for the Maxwell field to be (anti-)self dual by a proper choice of the parameters without spoiling either the bolt structure or the reality of the scalar field. Remarkably, this issue can be solved by  fixing the parameter of the Chern-Pontryagin density as $\Theta=\pm1/4$. For this particular choice, the Maxwell sector can be written as
\begin{align}\label{Maxwellinstanton}
 -\frac{1}{4}F^2 - \Theta\tilde{F}F =   -\frac{1}{4}\left(F^2\pm\tilde{F}F \right) = -\frac{1}{8}\left(F_{\mu\nu}\pm\tilde{F}_{\mu\nu} \right)\left(F^{\mu\nu}\pm\tilde{F}^{\mu\nu}\right)\,.
\end{align}
Indeed, the last equation vanishes identically for (anti-)self dual $U(1)$ connection, although the solution in Eq.~\eqref{sol1a} is not. This choice implies that the electromagnetic ground state of the theory is globally (anti-)self dual. Then, the renormalized Euclidean on-shell action for the Maxwell field becomes
\begin{align}
     - \frac{1}{8}\int\diff{^4x}\sqrt{|g|}\left(F_{\mu\nu}\pm\tilde{F}_{\mu\nu} \right)\left(F^{\mu\nu}\pm\tilde{F}^{\mu\nu}\right) =  \frac{4\pi^2 Q^2}{r_b^2\left(r_b^2+2b\right)}\,.
\end{align}

The renormalization of $I_{\rm bulk}$ including the gravitational sector in Eq.~\eqref{Ibulk} is rather nontrivial. This is related to the fact the solution is not asymptotically locally AdS. Indeed, the series of intrinsic boundary counterterms cannot be truncated as a consequence of dimensionality due to their falloff towards the asymptotic boundary. Thus, the standard prescription of intrinsic boundary counterterms is ambiguous here. We postpone a deeper study of this issue for the future. Nevertheless, there is a remarkable counterexample that occurs when the theory becomes conformally invariant, whose action is finite without any reference to intrinsic boundary counterterms. We provide the explicit computation of that case in Sec.~\ref{sec:conformal}. 

% Even though the Maxwell field in Eq.~\eqref{sol1} is not (anti-)self dual,\footnote{There exists a particular choice of the parameters that renders the Maxwell field (anti-)self dual, i.e. $\alpha = -\tfrac{2}{9}\pi G\Lambda$. In AdS, however, this condition is incompatible with reality of the scalar field and the existence of a bolt at the same time.} the counterterms 
% \begin{align}\label{CTsol1}
%   \Theta &= \pm\frac{1}{4}\,, & \zeta_1 &= -\frac{37}{96\pi G \ell}\,, & \zeta_2 &= -\frac{7\ell}{192\pi G}\,, 
% \end{align}
% renders the Euclidean on-shell action finite, giving explicitly
% \begin{align}
%     -I_E = \frac{\pi\,\beta_\tau}{24}\left[\frac{24\,Q^2}{r_b^4} - \frac{8r_b^4-64\ell^2r_b^2-27\ell^4}{32\pi G\ell^2} + \frac{\left(r_b^2+\ell^2\right)\left(r_b^2-3\ell^2\right)}{\alpha\ell^4} \right]\,.
% \end{align}

% \bc{Cristóbal: We should obtain the electric potential to compute the thermodynamics quantities, i.e., internal energy, entropy, and electric charge following Ref.~\cite{Chamblin:1999tk}.}

When $\gamma=-1$, the divergent piece of the Maxwell field in Eq.~\eqref{sol1a} can be eliminated by a proper choice of the parameters while keeping reality of the scalar field, metric regularity, and the asymptotically locally Euclidean AdS behavior; this choice is 
\begin{align}
    \alpha = \frac{4\pi G}{3\ell^2}\,.
\end{align}
Then, the solution to the field equations becomes
\begin{align}\label{EHkm1}
   f(r) &= -1 + \frac{b}{r^2} + \frac{r^2}{\ell^2}\,, & \phi(r) &= \frac{1}{r^2}\sqrt{\frac{3b\ell^2}{4\pi G}} \,, & A &= \frac{Q}{r^2}\,\tilde{\sigma}_3\,,
\end{align}
where $\tilde{\sigma}_i$ is obtained by setting $\sin\vartheta\to\sinh\vartheta$ and $\cos\vartheta\to\cosh\vartheta$ in Eq.~\eqref{MCformsu2}, according to the metric structure~\eqref{Eguchi-Hanson-ansatz} when $\gamma=-1$. Indeed, the Maxwell field is (anti-)self dual in this case, implying that its energy-momentum vanishes. Moreover, provided the choice of the parameter $\Theta=\pm1/4$, the Euclidean on-shell action for the Maxwell sector vanishes as well, setting the (anti-)self-dual configuration as the electromagnetic ground state of the system, as it was shown in Ref.~\cite{Araneda:2016iiy}.

\subsection{Taub-NUT}

There exists an additional gravitational instanton in Einstein gravity in presence of a conformally coupled scalar field. This solution was studied in Refs.~\cite{Bardoux:2013swa,Bhattacharya:2013hvm} and we compute some of their global properties here for the first time. To this end, we focus on the Taub-NUT metric based on the $U(1)$ fibration of constant curvature K\"ahler manifolds, namely, 
 \begin{align}\label{Taub-NUT-ansatz}
     \diff{s^2} = f(r)\left(\diff{\tau} + 2nB_{(\gamma)} \right)^2 + \frac{\diff{r^2}}{f(r)} + \left(r^2 -n^2\right)\diff{\Sigma_{(\gamma)}^2}\,,
  \end{align}
where, recall, $B_{(\gamma)}$ and $\diff{\Sigma_{(\gamma)}^2}$ have been defined in Eq.~\eqref{basemanifold}. The NUT charge $n$ sources the magnetic part of the Weyl tensor and it is usually interpreted as the gravitomagnetic mass since it produces a Lorentz-type force for test particles moving on geodesics over the Taub-NUT solution to GR~\cite{Lynden-Bell:1996dpw}. Moreover, it plays the role of a squashing parameter in the asymptotic region~\cite{Hartnoll:2005yc,Bobev:2016sap,Bueno:2018yzo} and it is related to the first Chern number at infinity~\cite{Hawking:1998ct}.  

The field equations~\eqref{eom} admit the following analytic solution~\cite{Bardoux:2013swa,Bhattacharya:2013hvm}
\begin{subequations}
\label{solcharmou}
\begin{align}
    f(r) &= -\frac{\Lambda\left(r^2-n^2 \right)}{3} + \frac{\left(\gamma+\frac{4}{3}\Lambda n^2\right)\left(r-\mu\right)^2}{r^2-n^2}\,, \\ 
    A &= \left[\frac{qr}{r^2-n^2} + \frac{p}{2n}\frac{r^2+n^2}{\left(r^2-n^2\right)} \right]\left(\diff{\tau} + 2n\cos\vartheta\diff{\varphi} \right)\,,\\
    \phi(r) &= \frac{1}{r-\mu}\sqrt{-\frac{\Lambda\left( \mu^2-n^2\right)}{6\alpha}} \,, 
\end{align}
\end{subequations}
where $\mu$, $q$ and $p$ are integration constants related through
\begin{align}\label{constraint_taubnut}
   \frac{27\alpha}{2\pi G}\left[\left(\mu^2-n^2 \right)\left(\gamma+\frac{4}{3}\Lambda n^2\right) + 4\pi G\left(q^2-p^2 \right) \right] + 3\Lambda\left(\mu^2-n^2 \right)\left(\gamma+\frac{4}{3}\Lambda n^2\right) = 0\,.
\end{align}
Reality of the scalar field when $\Lambda<0$ demands that either $n^2<\mu^2$ and $\alpha>0$ or $n^2>\mu^2$ and $\alpha<0$. The globally (anti-)self-dual condition on the Weyl tensor is achieved when
\begin{align}
    f(n) = 0 \;\;\;\;\; \mbox{and} \;\;\;\;\; f'(r)|_{r=n} = \frac{4\pi}{\beta_\tau}\,,
\end{align}
where prime denotes differentiation with respect to the radial coordinate and $\beta_\tau$ is the period of the Euclidean time. These conditions implies that $\mu=n$ and $\beta_\tau = 8\pi n$. Nevertheless, in this case the scalar field vanishes identically, the Maxwell field becomes (anti-)self dual when $p=q$, and the space becomes completely regular. The $U(1)$ gauge field of Eq.~\eqref{solcharmou} matches the one found in Ref.~\cite{Pope:1981jx} (see also~\cite{Boulton:2021wln}) at the (anti-)self-dual point $p=q$. Therefore, its energy-momentum vanishes identically. Indeed, the solution reduces to the standard Taub-NUT solution of Einstein-(A)dS gravity possessing a (anti-)self dual Weyl and Faraday tensors. 

In the case when $\gamma=1$, the Euler characteristic is $\chi(\mathcal{M}) = 1$. The Atiyah-Patodi-Singer index theorem for the Dirac operator $i\slashed{D}\equiv iE^\mu_a\gamma^a D_\mu$ for a manifold with boundary is given by~\cite{APS-eta}\footnote{Here, $E^\mu_a$ is the inverse of the tetrad field $e^a_\mu$ defined as $g_{\mu\nu}=e^a_\mu e^b_\nu \delta_{ab}$, i.e. $e^a_\mu E^\mu_b = \delta^a_b$ and $e^a_\mu E^\nu_a = \delta_\mu^\nu$. }
\begin{align}\notag
    n_+ - n_- &= \frac{1}{24}\Bigg[ \frac{1}{32\pi^2} \int_{\mathcal{M}} d^4x\,\sqrt{g}\, \varepsilon_{\mu\nu \lambda\rho }R_{\sigma\tau}^{\lambda\rho}R^{\sigma\tau\mu\nu } - \frac{1}{4\pi^2}\int_{\partial\mathcal{M}}   \diff{^3x}\sqrt{h}\;n_\mu\varepsilon^{\mu\nu\lambda\rho}K^\sigma_\nu\nabla_\lambda K_{\rho\sigma}\Bigg] \\
    \label{Diracindex}
    &\quad -\frac{1}{2}\left[\eta_D\left(\partial\mathcal{M} \right) + h\left(\partial\mathcal{M}\right) \right]\,,
\end{align}
where $n_\pm$ are the number of positive and negative chiral spinors statisfying the boundary conditions defined in Ref.~\cite{APS-eta},  $\eta_D\left(\partial\mathcal{M}\right)$ measures the difference between positive and negative eigenvalues of the tangential components of the Dirac operator on $\partial\mathcal{M}$, and $h\left(\partial\mathcal{M}\right)$ is the dimension of the space of functions harmonic under $i\slashed{D}$. Furthermore, whenever edges-conical singularities are present, the Euler characteristic and the Dirac index receive an additional correction proposed in~\cite{Atiyah:2011fn,atiyah_lebrun_2013}. Nevertheless, the solutions presented here are absent of edge-conical singularities once the right periodicity of the Euclidean time is imposed [cf. Eq.~\eqref{betasolEH}]. Therefore, their contribution to these topological invariants vanishes. 

For vanishing cosmological constant, the value $\eta_D=-1/6$ was found in Refs.~\cite{Eguchi:1977iu,Pope:1978zx,Pope:1981jx} for Taub-NUT using the Hitchin's formula~\cite{HITCHIN19741}. On the other hand, in presence of negative cosmological constant, we find 
\begin{align}
    \frac{1}{2}\left[\eta_D\left(\partial\mathcal{M} \right) + h\left(\partial\mathcal{M}\right) \right] = \frac{1}{12} - \frac{2n^2}{3\ell^2}\left(1-\frac{2n^2}{\ell^2} \right)\,.
\end{align}
Thus, performing the first two integrals of Eq.~\eqref{Diracindex} that define the Chern-Pontryagin index and the Chern-Simons form, respectively, we obtain
\begin{align}
    n_+ - n_- &= \frac{1}{24}\left[2-\frac{16n^2}{\ell^2}\left(1-\frac{2n^2}{\ell^2}\right)\right] - \frac{1}{12} + \frac{2n^2}{3\ell^2}\left(1-\frac{2n^2}{\ell^2} \right) = 0\,.
\end{align}
This implies that there is no asymmetry between Dirac spinors with different chirality; this is contrary to the Hermitian operator associated to Rarita-Schwinger fields~\cite{Eguchi:1977iu}. 

Although the globally (anti-)self-dual case holds only for a constant scalar field, one can still demand a weaker condition: asymptotically (anti-)self-dual configurations, as in the Taub-Bolt instanton. In this case, the geometry has a bolt at $r=r_b$ defined by $f(r_b)=0$ with $r_b>n$. This condition produces a $2$-dimensional set of fixed points and it relates the parameters as
\begin{align}
    \mu = \mu_{\rm bolt} \equiv r_b \pm \left(r_b^2-n^2 \right)\sqrt{\frac{\Lambda}{3\left(\gamma+\frac{4}{3}\Lambda n^2\right)}} \,.
\end{align}
Additionally, the absence of conical singularities at the bolt is obtained by imposing periodicity on the Euclidean time coordinate, i.e., $\tau\sim\tau + \beta_\tau$, with
\begin{align}
    \beta_\tau = - \frac{6\pi}{2\Lambda r_b \pm \sqrt{3\Lambda\left(\gamma+\frac{4}{3}\Lambda n^2\right)}}\,,
\end{align}
where the $2\Lambda r_b \pm \sqrt{3\Lambda\left(\gamma+\frac{4}{3}\Lambda n^2\right)}<0$ such that $\beta_\tau$ is positive. For $\gamma=1$ the Euler characteristic of this solution is $\chi\left(\mathcal{M}\right) = 2$. Moreover, the existence of a bolt alongside reality of the Euclidean time's period constrains the NUT charge according to the condition $\Lambda<0$ and
\begin{align}
    \frac{4\Lambda}{3}\left(r_b^2-n^2\right)<\gamma<-\frac{4\Lambda n^2}{3}\,.
\end{align}

Since the Taub-Bolt solution is asymptotically locally AdS, a well-posed variational and renormalized action principle can be obtained by adding boundary terms of AdS gravity to the action~\eqref{Ibulk}; these are~\cite{Emparan:1999pm,Balasubramanian:1999re}
\begin{align}
\label{IGHY}
    I_{\rm GHY} &= 2\int_{\partial\mathcal{M}}\diff{^3x}\sqrt{|h|}\left[\kappa-\frac{\phi^2}{12}\right]K\,, \\
    \label{Ict}
    I_{\rm ct} &= \int_{\partial\mathcal{M}}\diff{^3x}\sqrt{|h|}\left(\zeta_1 + \zeta_2\mathcal{R} + \zeta_3\phi^2\mathcal{R} + \zeta_4 h^{\mu\nu}\nabla_\mu\phi\nabla_\nu\phi  + \zeta_5\mathcal{R}^2 + \ldots \right)\,,
\end{align}
where the induced metric on the boundary is defined as  $h_{\mu\nu}=g_{\mu\nu}-n_\mu n_\nu$, its determinant is $h=\det h_{\mu\nu}$, and $n^\mu$ is the space-like normal unit vector defining radial foliation. The extrinsic curvature is  $K_{\mu\nu}=h^{\lambda}{}_\mu\nabla_\lambda n_\nu$, while $K=h^{\mu\nu}K_{\mu\nu}$ denotes its trace. The intrinsic curvature $\mathcal{R}^{\mu\nu}_{\lambda\rho}$ is defined through the Gauss-Codazzi equation 
\begin{align}
    \mathcal{R}^{\mu\nu}_{\lambda\rho} = h^{\mu}{}_{\alpha} h^{\nu}{}_{\beta} h^{\gamma}{}_{\lambda} h^{\delta}{}_{\rho} R^{\alpha\beta}_{\gamma\delta} + K^\mu_{\lambda} K^{\nu}_{\rho} - K^\mu_{\rho} K^{\nu}_{\lambda}  \,,
\end{align}
the intrinsic Ricci tensor is denoted by $\mathcal{R}^\mu_\nu=\mathcal{R}^{\mu\lambda}_{\nu\lambda}$ and $\mathcal{R} \equiv \mathcal{R}^{\mu\nu}_{\mu\nu}$ is the intrinsic Ricci scalar. We consider the full Euclidean on-shell action to be $-I_E = I_{\rm bulk} + I_{\rm GHY} + I_{\rm ct}$, where $I_{\rm bulk}$ is given in Eq.~\eqref{Ibulk}. The role of the boundary term $I_{\rm GHY}$ is two-fold: first, it is included to guarantee a well-posed variational principle and, second, it cancels the boundary contribution coming from the bulk action after Weyl rescalings. In other words, it renders the scalar-tensor sector of the theory invariant rather than quasi-invariant under conformal transformations.\footnote{We thank to G.~Anastasiou for pointing this out to us.} Additionally, the term $I_{\rm ct}$ are standard intrinsic boundary counterterms that render the Euclidean on-shell action finite in presence of AdS asymptotics. Nevertheless, only the first two terms in Eq.~\eqref{Ict} contribute to the Euclidean action of the Taub-Bolt-AdS solution, since the remaining ones decay sufficiently fast asymptotically. 

Renormalization of the Euclidean on-shell action demands that $\zeta_1=-4\kappa/\ell$ and $\zeta_2=-\kappa\ell$. Then, for $\gamma=1$, we obtain
\begin{align}\notag
    -I_E &= \frac{\beta_\tau\,\pi}{3\alpha\left(1+\frac{4}{3}\Lambda n^2\right)}\Bigg[\frac{2r_b\Lambda^3}{27}\left(5n^2+3r_b^2\right)\left(3n^2-r_b^2\right) + \frac{r_b\Lambda^2}{9}\left[96\alpha\kappa n^2 \left(r_b^2+n^2\right) + 21n^2 - r_b^2\right] \\
    \notag
    &\quad + \Lambda\left\{r_b+\alpha\left[8\kappa r_b\left(5n^2+r_b^2\right) + \frac{8n^2r_b(p-q)^2 }{\left(r_b+n \right)^2}\right]\right\} + \alpha\left[24\kappa r_b + \frac{6r_b(p-q)^2}{(r_b+n)^2}\right] \\
    \notag
    &\quad \pm\sqrt{\frac{\Lambda}{3\left(1+\frac{4}{3}\Lambda n^2\right)}}\Bigg\{1+\Lambda\left(r_b^2+3n^2\right) + \frac{\Lambda^2}{3}\left(8n^4+10n^2r_b^2-2r_b^4\right)\\
    &\quad +\frac{2\Lambda^3}{27}\left(3n^2-r_b^2\right)\left(3n^4+12n^2r_b^2+r_b^4\right)
    +24\kappa\alpha\left(r_b^2-n^2\right)\left(1+\frac{4}{3}\Lambda n^2\right)^2\Bigg\}\Bigg]\,.
\end{align}
%\pc{Marcelo: Agree! I used $\zeta_1=-\frac{\sqrt{2}}{4\ell \pi G}=\frac{4\Lambda\kappa}{\sqrt{-3\Lambda}} \,,\,\,\zeta_2=-\frac{\sqrt{2}\ell}{32\pi G}=-\frac{3\kappa}{\sqrt{-3\Lambda}}$ in \eqref{Ict} to renormalize the on-shell action. I used $\Lambda=-6/\ell^2$ .}
To first order in the saddle-point approximation, the Euclidean on-shell action is identified with the free energy of the system through $\mathcal{F} = \beta_\tau^{-1}I_E$. Indeed, it is well-known that some gravitational systems can develop a Hawking-Page phase transition in AdS from the ground state to a large black hole configuration at some critical temperature~\cite{Hawking:1982dh}. In particular, the case of Taub-NUT/Bolt-AdS in Einstein gravity has been explored in Refs.~\cite{Johnson:2014xza,Johnson:2014pwa}. It might be interesting to analyze whether the phase structure found here allows for the system to develop a Hawking-Page phase transition with a nontrivial scalar field. We postpone a deeper study of this point for the future. 

% Finally, the global properties of the Taub-Bolt solution found here are labeled by the Hirzebruch signature and the Euler characteristic. When $\gamma=1$, these topological invariants for the Taub-Bolt solution found here are respectively given by
% \begin{align}
%     \tau\left(\mathcal{M}\right) = -\frac{4n\beta_\tau\Lambda}{27\pi}\left[\Lambda\left(2r_b^2+n^2\right) \pm 2r_b\sqrt{3\Lambda\left(1+ \frac{4\Lambda n^2}{3}\right)}\right] \;\;\;\;\; \mbox{and} \;\;\;\;\; \chi\left(\mathcal{M}\right) = 2\,.
% \end{align}

In the next section, we show how a suitable choice of the coupling constant associated to higher-curvature corrections can eliminate the divergent piece of the Maxwell field in Eq.~\eqref{sol1a} while keeping a positive curvature K\"ahler base manifold and negative cosmological constant. 

\section{Higher-curvature corrections\label{sec:higher-curvature}}

It is well-known that higher-curvature corrections might arise at the low-energy limit of ultraviolet completion gravitational theories~\cite{Zwiebach:1985uq,Deser:1986xr,Duff:1986pq,Cano:2021rey}. Indeed, the latter are known to renormalize the Einstein-Hilbert action perturbatively around the Minkowski background~\cite{Stelle:1976gc}. Additionally, they become important in early periods of the Universe~\cite{Starobinsky:1980te} and when renormalization of the energy-momentum tensor in curved spacetimes is considered~\cite{Davies:1977ze,Birrell:1982ix}. 

In this section, we study how higher-curvature corrections modify the gravitational instantons presented in the previous section. In particular, we focus on quadratic terms in the curvature given by
\begin{equation}\label{Ihc}
     I_{\rm hc}\left[g_{\mu\nu},\phi \right] = \int_{\mathcal{M}}\diff{^4x}\sqrt{|g|}\left[c_1 R^2+c_2 \phi^{-4}S^2\right]\,,
\end{equation}
such that the total action becomes $-I_E=I_{\rm bulk} + I_{\rm GHY} + I_{\rm ct} + I_{\rm hc}$. Here, $c_1$ and $c_2$ are arbitrary parameters and we have introduced the trace of the tensor (see~\cite{Oliva:2011np} for details)
\begin{equation}\label{Stensor}
    S^{\mu\nu}_{\,\lambda\rho}=\phi^2R^{\mu\nu}_{\,\lambda\rho}-4\phi\delta^{[\mu}_{[\lambda}\nabla^{\nu]}\nabla_{\rho]}\phi+8\delta^{[\mu}_{[\lambda}\nabla^{\nu]}\phi\nabla_{\rho]}\phi-\delta^{\mu\nu}_{\lambda\rho}\nabla_\alpha\phi\nabla^\alpha\phi\,,
\end{equation}
which is explicitly given by
\begin{equation}
    S = S^{\mu\nu}_{\mu\nu} = \phi^2 R-6\phi\square\phi \ .
\end{equation}
Interestingly enough, as $g_{\mu\nu}\to\Omega^2(x)g_{\mu\nu}$ and $\phi\to\Omega^{-1}(x)\phi$, the tensor $S^{\mu\nu}_{\lambda\rho}$ in Eq.~\eqref{Stensor} transforms covariantly under Weyl rescaling, namely,~\cite{Oliva:2011np}
\begin{align}
    S^{\mu\nu}_{\lambda\rho}\to  \Omega^{-4}S^{\mu\nu}_{\lambda\rho}\,.
\end{align}

The bulk action~\eqref{Ibulk} modified by the higher-curvature terms in $I_{\rm hc}$, leads to field equations for the metric and scalar field which are similar to those in Eq.~\eqref{eom}. However, there appear additional higher-derivative corrections induced by quadratic pieces in the Ricci scalar. Specifically, the higher-curvature terms in Eq.~\eqref{Ihc} yield
\begin{subequations}\label{eomhc}
\begin{align}
 \mathcal{E}_{\mu\nu} &= 2\kappa\left(G_{\mu\nu} + \Lambda g_{\mu\nu}\right) - T_{\mu\nu}- c_1\,T_{\mu\nu}^{(R^2)} - c_2\,T_{\mu\nu}^{(S^2)}= 0,\\
\mathcal{E} &= \Box\phi - \frac{1}{6}R\phi - 4\alpha\phi^3+4c_2\left[-S^2\phi^{-5}+\phi^{-3}RS-3S\phi^{-4}\square\phi-3\square\left(\phi^{-3}S\right)\right]=0,
\end{align}
\end{subequations}
where we have defined 
\begin{align}
    T_{\mu\nu}^{(R^2)}&=2g_{\mu\nu}\square R-2\nabla_\mu\nabla_\nu R+2RR_{\mu\nu}-\frac{1}{2}g_{\mu\nu}R^2, \\
    \notag
    T_{\mu\nu}^{(S^2)}&=\left(2R_{\mu\nu}-2\nabla_\mu\nabla_\nu+2g_{\mu\nu}\square\right)\left(\phi^{-2}S\right)+12\nabla_{(\mu}\left(\phi^{-3}S\right)\nabla_{\nu)}\phi\\
    &\quad -6g_{\mu\nu}\nabla^\alpha\left(\phi^{-3}S\nabla_\alpha\phi\right)-\frac{1}{2}g_{\mu\nu}\phi^{-4}S^2.
    \end{align}

To solve the field equations in this case, we consider the Eguchi-Hanson-like ansatz in Eq.~\eqref{Eguchi-Hanson-ansatz}. Then, Eqs.~\eqref{eomhc} admit the following metric function, scalar and Maxwell field as solutions 
\begin{align}
    f(r) &= \gamma + \frac{b}{r^2}-\frac{\Lambda}{6}r^2\,, \\
    \phi(r) &=\frac{1}{r^2}\sqrt{\frac{b}{\alpha}}\,, \\
    A &= \left[\frac{q}{r^2}-\frac{b\left[72\kappa \alpha +\Lambda +\left(
c_{1}+c_{2}\right) 288\Lambda \alpha\right]}{576\,q\,\alpha}\;r^2\right]2\sigma_3\,,
\end{align}
%\pc{Marcelo: I obtained the same solution with $%
%\gamma =+1,-1,0$, but the field equations fix a different value for the
%constant which multiply the $r^{2}$ term in the gauge field:%
%\begin{equation*}
%A=\left[ \frac{q}{r^{2}}-\frac{b\left[ 72\kappa \alpha +\Lambda +\left(
%c_{1}+c_{2}\right) 288\Lambda \alpha \right] }{576q\alpha }\right] \left(
%d\tau +\cos \theta d\varphi \right) \ .
%\end{equation*}%
%The term proportional to $c_{1}+c_{2}$ is in fact the problem. Please check
%that! Thanks.}
where $\alpha\neq0$ and $q\neq0$.  Thus, we notice that the introduction of these particular higher-curvature corrections induces a shift in the Maxwell field. Nevertheless, an interesting consequence of this modification is that, in contrast to the solution in Eq.~\eqref{sol1}, an asymptotically locally Euclidean AdS instanton possessing a compact K\"ahler manifold appears when $\Lambda<0$, whenever the higher-curvature couplings satisfy the relation
\begin{align}
 c_1 + c_2 = -\frac{1}{4}\left(\frac{\kappa}{\Lambda}+\frac{1}{72\alpha}\right).
\end{align}
With this choice, the scalar field is nontrivial and the Maxwell field becomes (anti-)self dual. Thus, the latter represents the electromagnetic ground state of the theory as long as the $U(1)$ Pontryagin coupling is chosen as $\Theta=\pm1/4$.

The case of Taub-NUT metric in quadratic gravity with conformally coupled scalar field was studied in Ref.~\cite{Cisterna:2021xxq}. Indeed, the authors found a smooth embedding of solution~\cite{Bardoux:2013swa} in theories with higher-curvature corrections of the type described above. The embedding is a consequence of the conformal invariance of the terms involving the $S^{\mu\nu}_{\lambda\rho}$ tensor and the fact that the quadratic contribution $R^2$ naturally supports solutions with constant Ricci scalar in four dimensions.  Moreover, the authors showed that a double-Wick rotation allows one to obtain a Lorentzian traversable wormhole; similar to the ones presented in~\cite{Anabalon:2018rzq,Anabalon:2020loe}. 

In the next section, we study a conformally invariant scalar-tensor theory of the Brans-Dicke type. This case is interesting because the conformal symmetry allows one to renormalize the Euclidean on-shell action and conserved charges naturally without any reference to intrinsic boundary counterterms. This feature bears close resemblance to the pure metric formulation of conformal gravity~\cite{Grumiller:2013mxa,Anastasiou:2020mik,Anastasiou:2021tlv,Corral:2021xsu}. We show this explicitly for a novel set of gravitational instantons we present next.

\section{Conformally invariant scalar-tensor gravity\label{sec:conformal}}

There is an interesting point in the parameter space of Brans-Dicke gravity that leads to a conformally invariant theory. The latter can be obtained when the Einstein sector decouples from of the action~\eqref{Ibulk}, producing a second-order set of field equations given by $\tilde{\mathcal{E}}_{\mu\nu}=0$, $\mathcal{E}^\mu=0$, and $\mathcal{E}=0$, where the last two terms are defined in Eqs.~\eqref{eoma} and~\eqref{eomp}, respectively, while the former is given by
\begin{align}\notag
   \tilde{\mathcal{E}}_{\mu\nu} &= G_{\mu\nu} + \frac{6}{\phi^2}\left(\nabla_\mu\phi\nabla_\nu\phi - \frac{1}{2}g_{\mu\nu}\nabla_\lambda\phi\nabla^\lambda\phi - \alpha g_{\mu\nu}\phi^4 + F_{\mu\lambda}F_{\nu}{}^\lambda - \frac{1}{4}g_{\mu\nu}F^2\right) + g_{\mu\lambda}\delta_{\nu\tau}^{\lambda\rho}\nabla^\tau\nabla_\rho\phi^2 \,.
\end{align}
Interestingly enough, this case leads to novel analytical charged gravitational instantons with conformally coupled scalar fields. Moreover, their Euclidean on-shell action and Noether-Wald charges are finite in spite of their nontrivial asymptotics. These solutions are discussed next.    

\subsection{Eguchi-Hanson}

The field equations~\eqref{eom} admit a different solution of the Eguchi-Hanson family in the conformal point of Brans-Dicke gravity. We focus on the choice $\gamma=1$ for the sake of simplicity. The metric function in Eq.~\eqref{Eguchi-Hanson-ansatz}, as well as Maxwell and scalar fields are given by
\begin{subequations}\label{solutionMarcelo}
\begin{align}\label{fsolmarcelo}
    f(r) &= k + \frac{r^2}{\ell^2} + \frac{b}{r^2} + \frac{\ell^2\left( k-1\right)\left[9b-\ell^2\left(k-1 \right)^2\right]}{54\,r^4}\,,\\
    A &= \left[\frac{q}{r^2} + \frac{\left[3b-\ell^2\left(k^2-1 \right)\right]r^2}{288\alpha\ell^2 q} \right]2\sigma_3 \,,\\
    \phi(r) &= \frac{1}{r^2+\tfrac{\ell^2}{3}\left(k-1 \right)} \sqrt{\frac{b-\tfrac{\ell^2}{3}\left(k^2-1\right)}{\alpha}}\,,
\end{align}
\end{subequations}
respectively, where $k$, $\ell$, $b$, and $q$ are integration constants. Notice that this solution only exists for $\alpha\neq0$. Equation~\eqref{fsolmarcelo} suggests that $k$ behaves as an effective curvature of the K\"ahler base manifold while $\ell^2$ can be interpreted as the inverse of a negative cosmological constant. When $k=1$, one recovers the solution in Eq.~\eqref{sol1} provided a proper identification between the parameters where, recall, $\gamma=1$ has been assumed. This implies that Eq.~\eqref{fsolmarcelo} includes the previous solution as a particular case. Additionally, the cases $k=0$ and $k=-1$ are discussed in Table~\ref{tabla}. These, however, are not globally (anti-)self dual but, rather, asymptotically (anti-)self dual.
\begin{center}
\begin{table}[!ht]
\small
\addtolength{\tabcolsep}{-5pt}
\centering
    \begin{tabular}{ |c|c|c| }
   \hline
    & Case when $k=0$ & Case when $k=-1$ \\
    \hline
   $f(r)$ & $\frac{r^2}{\ell^2} + \frac{b}{r^2} - \frac{\ell^2\left(9b-\ell^2 \right)}{54\,r^4}$ & $-1 + \frac{r^2}{\ell^2} + \frac{b}{r^2}    - \frac{\ell^2\left(9b-4\ell^2 \right)}{27\,r^4}$ \\
    \hline
    $\phi(r)$ & $\frac{1}{r^2-\ell^2/3}\sqrt{\frac{3b+\ell^2 }{3\alpha}}$ & $\frac{1}{r^2-2\ell^2/3}\sqrt{\frac{b}{\alpha}}$\\
    \hline
    $A$ & $\left(\frac{q}{r^2} + \frac{\left(3b+\ell^2 \right)r^2}{288\,q\,\alpha\,\ell^2} \right)2\sigma_3$ & $\left(\frac{q}{r^2} + \frac{b\,r^2}{96\,q\,\alpha\,\ell^2} \right)2\sigma_3$ \\
    \hline
    $\beta_\tau$ & $\frac{2\pi\,r_b^2\,\ell^2\,\left(r_b^2-\ell^2/6\right)}{r_b^6-r_b^4\ell^2/4-\ell^6/108}$ & $\frac{2\pi r_b^2\ell^2\left(r_b^2-\ell^2/3\right)}{r_b^6-r_b^4\ell^2 + r_b^2\ell^4/3-2\ell^2/27}$ \\
    \hline
   $-I_{\rm bulk}$ & $\frac{4\pi\beta_\tau q^2}{r_b^4} + \frac{\pi\beta_\tau}{24}\frac{\left(3b+\ell^2\right)\left(7\ell^4-45\ell^2r_b^2 + 108r_b^4-15b\ell^2+81b r_b^2\right)}{\alpha\left(3r_b^2-\ell^2\right)^3}$ & $\frac{4\pi\beta_\tau q^2}{r_b^4} + \frac{27\pi\beta_\tau b}{2\alpha}\frac{\left[b\left(3r_b^2/4-5\ell^2/18\right) + \left(r_b^2-2\ell^2/3\right)^2 \right]}{\left(3r_b^2-2\ell^2\right)^3} $ \\ 
    \hline
    $-I_{\rm GHY}$ & $\frac{\pi\beta_\tau}{9\alpha\ell^2}\left(3b+\ell^2\right)$ & $\frac{\pi\beta_\tau b}{3\alpha\ell^2}$ \\
    \hline
    \end{tabular}
    \caption{\label{tabla}Particular cases of the solution Eq.~\eqref{solutionMarcelo} for different values of $k$. When $k=1$, the solution reduces to the one presented in Eq.~\eqref{sol1}, modulo a redefinition of the parameters. In all cases, the scalar field remains finite as $r\to r_b$, where $r_b$ is defined through $f(r_b)=0$. Moreover, the spacetime is absent of conical singularities if the period of the Euclidean time is $\beta_\tau$. The Weyl tensor in these cases is not globally (anti-)self dual but, rather, asymptotically (anti-)self dual.}
\end{table}
\end{center}

There is another interesting possibility, though: take $k\to1$ as $\ell\to\infty$ keeping $\ell^2(k-1)$ fixed and equal to $-6a^4/b$. This yields
\begin{align}\label{sol2}
    f(r) &= 1 - \frac{a^4}{r^4} + \frac{b}{r^2}\,, & \phi(r) &= \frac{1}{r^2-2a^4/b}\sqrt{\frac{4a^4+b^2}{\alpha b}}\,, & A= \frac{q}{r^2}\,\sigma_3\,.
\end{align}
Indeed, the Weyl tensor of this solution is globally (anti-)self dual and it represents a particular case of the solution found in Ref.~\cite{Corral:2021xsu} but supported by nontrivial scalar and Maxwell fields. The scalar field is finite at $r=r_b$ defined through $f(r_b)=0$. The Maxwell field, in turn, is globally (anti-)self dual. Thus, even though its field strength is nonvanishing, their energy-momentum tensor is zero. Additionally, since the $U(1)$ Pontryagin density is conformally invariant, one can still keep it at the action level and fix the $\Theta$ parameter such that the (anti-)self dual Maxwell field becomes the ground state of the theory, similar to the case in Eq.~\eqref{EHkm1}.

The solution is endowed with a conical singularity at $r=r_b$ for the standard period of the Euler's angles. Nevertheless, it can be removed by demanding that the period of the Euclidean time is given by
\begin{align}
    \beta_\tau = \frac{4\pi r_b^2}{2r_b^2+b}\,.
\end{align}
As usual, the Hawking temperature can be obtained from the standard relation $T=\beta_\tau^{-1}$. Remarkably, as we pointed out, conformal invariance protects the bulk action against divergences while the generalized Gibbons--Hawking--York term vanishes; they are
\begin{align}\label{IEHconformal}
   - I_{\rm bulk} =  \frac{\pi^2}{6\alpha}\left(2-\frac{b}{r_b^2} \right) %- \frac{16\pi^2 q^2}{r_b^2\left(2r_b^2+b\right)}\,.
   \;\;\;\;\; \mbox{and} \;\;\;\;\; I_{\rm GHY} = 0\,.
\end{align}
% Here, renormalizability of the Maxwell sector does not fix the $\Theta$-parameter as in the case when $\kappa\neq0$. However, we keep its value such that the relation in Eq.~\eqref{Maxwellinstanton} holds. 
Indeed, we check explicitly that the value of Eq.~\eqref{IEHconformal} can be obtained from the general solution~\eqref{solutionMarcelo} by taking the limit $k\to1$ as $\ell\to\infty$, keeping fixed $\ell^2(k-1)=-6a^4/b$. On the other hand, the Hirzebruch signature and Euler characteristic of this solution are~\cite{Corral:2021xsu}
\begin{align}
    \tau(\mathcal{M}) = -\left(1 + \frac{2b}{3r_b^2} + \frac{b^2}{6r_b^4}\right)\left(1+\frac{b}{2r_b^2} \right)^{-1}\;\;\;\;\; \mbox{and} \;\;\;\;\; \chi(\mathcal{M}) = 2\,,
\end{align}
where we have used the fact that, asymptotically, the boundary metric is topologically $\mathbb{RP}^3$ and therefore the $\eta_S$-invariant vanishes~\cite{Hanson:1978uv}. This can be checked explicitly using the Hitchin's formula~\cite{HITCHIN19741}. In the limit $b\to0$, the scalar field vanishes and the solution reduces to the charged version of the (anti-)self dual Eguchi-Hanson instanton in Einstein gravity. When $a\to0$, in turn, the scalar profile is $\phi(r)=\frac{1}{r^2}\sqrt{\frac{b}{\alpha}}$ and the topological indices become $\tau(\mathcal{M}) = -1$ and $\chi(\mathcal{M}) = 2$. Thus, we conclude that their global properties are equivalent to the original Eguchi-Hanson instanton~\cite{Eguchi:1978xp,Eguchi:1978gw}. Nevertheless, the solution~\eqref{sol2} is endowed with a nontrivial scalar and Maxwell fields, and it is manifestly locally inequivalent to~\cite{Eguchi:1978xp,Eguchi:1978gw} as it can be seen from Eq.~\eqref{sol2} when $a\to0$.

\subsection{Taub-NUT}

Besides the (anti-)self dual Eguchi-Hanson instanton in the conformally invariant case, there is an additional Taub-NUT/Bolt-AdS solution with a weakened AdS asymptotics. This can be obtained by considering the line element~\eqref{Taub-NUT-ansatz} as a metric ansatz and noticing that field equations in the conformally invariant case are solved by
\begin{subequations}\label{solnutconformal}
\begin{align}
    f(r) &= \frac{\left(r-\mu\right)^2}{r^2-n^2}  +\frac{b\left(r-\mu\right)^3}{r^2-n^2} - \frac{\lambda\left(r-\mu\right)^3\left(r+3\mu\right)}{3\left(r^2-n^2\right)} + \frac{2\alpha\phi_0^2\left(6r^2+3\mu^2-8\mu r - n^2 \right)}{r^2-n^2}\,, \\
    A &= \left[\frac{qr}{r^2-n^2} + \frac{p}{2n}\frac{r^2+n^2}{\left(r^2-n^2\right)} \right]\left(\diff{\tau} + 2n\cos\vartheta\diff{\varphi} \right)\,,\\
    \phi(r) &= \frac{\phi_0}{r-\mu}\,,
\end{align}
\end{subequations}
where $\mu$, $b$, $\lambda$, $q$, $p$, $\phi_0$ are integration constants, related through
\begin{align}
    6\alpha\phi_0^4 + \left(1+\lambda\mu^2 - b\mu + \frac{\lambda n^2}{3}\right)\phi_0^2 - 3\left(q^2-p^2 \right) = 0\,.
\end{align}
In contrast to the globally (anti-)self dual Eguchi-Hanson instanton in the conformally invariant case, $\alpha$ is a free parameter possessing a smooth limit when $\alpha\to0$. The local asymptotics of the metric function in Eq.~\eqref{solnutconformal} exhibits the usual weakened AdS behaviour through the low decaying mode present in theories with conformal invariance; that is, as $r\to\infty$, the metric function behaves
\begin{align}\notag
    f(r) &= -\frac{\lambda\, r^2}{3} + b\,r + 1-3\mu b-\frac{\lambda}{3}\left(n^2-6\mu^2\right) + 12\alpha\phi_0 \\ 
    \label{asympsolnutconformal}
    &\quad - \frac{2\mu+16\alpha\mu\phi_0+\tfrac{8\lambda\mu^3}{3}-bn^2-3b\mu^2}{r} + \mathcal{O}(r^{-2})\,.
\end{align}

%\bc{Cristóbal: We need to study NUTs and bolts of this metric and check regularity of scalar fields in both cases at the fixed points. Moreover, for the Taub-NUT case, we need to obtain the Euclidean on-shell action, Chern-Pontryagin index, and the Euler characteristic. As in the Eguchi-Hanson case, I suspect that this Taub-NUT solution has finite Euclidean on-shell action in absence of the Einstein term. Please, check this.}

When the NUT condition is imposed, recall $f(n)=0$, the solution becomes (anti-)self dual and the following conditions on the parameters arise
\begin{align}\label{bandphifornutconformal}
    b = -\frac{2\left(1 + \frac{4}{3}\lambda n^2 \right)}{\mu-3n} \;\;\;\;\; \mbox{and} \;\;\;\;\; \phi_0^2 = -\frac{(\mu-n)\left[3+\lambda\left(\mu-n\right)^2\right]}{6\alpha\left(\mu-3n\right)}\,.
\end{align}
Reality of the scalar field imposes restrictions on the parameter space. This, in turn, requires that the mass of the Taub-NUT solution must be bounded from below. We study these properties in Sec.~\ref{sec:noether-wald}. Additionally, we notice that the absence of conical singularities implies that the period of the Euclidean time must be $\beta_\tau=8\pi n$.

The Euclidean on-shell action can be obtained by imposing the same condition on the $\Theta$-parameter in the Maxwell sector, recall, $\Theta=\pm1/4$, such that it can be written as in Eq.~\eqref{Maxwellinstanton}. This choice renormalize the infrared divergence at $r=n$. The Gibbons-Hawking-York term, on the other hand, is nontrivial and it contributes to the total Euclidean on-shell action that is given by  
\begin{align}\notag
    -I_E &= 4\pi^2\left(p-q\right)^2 - \frac{8\pi^2 n}{27\alpha\left(\mu-3n\right)^2}\bigg[2\left(\mu^2-3\mu n + n^2 \right)\left(\mu-n\right)^3\lambda^2 \\
    &\quad + 3\left(2\mu^2-5\mu n + n^2 \right)\left(\mu-n\right)\lambda + 9n \bigg].
\end{align}
The Dirac index and Euler characteristic for the Taub-NUT case in the conformal case are respectively given by
\begin{align}\label{indexnutconf}
    n_+-n_- = 0  \;\;\;\;\; \mbox{and} \;\;\;\;\; \chi(\mathcal{M}) = 1\,.
\end{align}
Thus, the (anti-)self-dual Taub-NUT solution~\eqref{solnutconformal} does not contribute to the axial anomaly, similar to what happens for the Taub-NUT solution of general relativity~\cite{Eguchi:1977iu}. Therefore, we conclude that both global and local properties of this solution are manifestly different from the ones presented in Ref.~\cite{Bardoux:2013swa}. Their dissimilarity with respect to the Eguchi-Hanson and Taub-Bolt instantons in the Euler characteristic shows that this solution has one boundary less, as it can be seen from its zero-dimensional set of fixed points: a NUT.

The Taub-Bolt case, on the other hand, is obtained by demanding that the set of fixed points of the Killing vector field $\xi=\partial_\tau$ are located at the codimension-2 hypersurface $r=r_b>n$, namely, $f(r_b)=0$. This condition, in turn, imposes that
\begin{align}\label{phi0boltconf}
   \phi_0^2 = \phi_{0\,{\rm bolt}}^2 \equiv \frac{(\mu-r_b)^2\left[1-b(\mu-r_b) + \frac{\lambda}{3}\left(3\mu^2-2\mu r_b - r_b^2\right) \right]}{2\alpha\left[n^2+8\mu r_b - 3\left(\mu^2+2r_b^2\right)\right]}\,.
\end{align}
The Weyl tensor associated to this Taub-Bolt solution is asymptotically locally (anti-)self dual. The absence of conical singularities can be achieved by demanding that the period of the Euclidean time is 
\begin{align}\notag
    \beta_\tau &= 12\pi\left(r_b^2-n^2\right)\left[n^2+8\mu r_b - 3\left(\mu^2+2r_b^2\right)\right]\Big\{\left(\mu-r_b\right)^2\big[4\lambda\left(3r_b^3-n^2\left[r_b+2\mu\right] \right) \\
    &\quad + 3b\left(3n^2-6r_b^2-\mu^2+4\mu r_b\right) - 6\left(\mu^2+n^2-2r_b\mu \right)\left(\mu-r_b\right)^{-1}\big] \Big\}^{-1}\,.
\end{align}
This condition implies that the metric is geodesically complete. Indeed, one can check that for the bolt condition~\eqref{phi0boltconf}, both the scalar and the Maxwell fields are finite at $r=r_b$. Thus, we conclude that this solution represents an asymptotically locally (anti-)self dual instanton with a bolt, possessing a Hawking temperature given by $T=\beta_\tau^{-1}$.

\subsection{Noether-Wald charges\label{sec:noether-wald}}

In conformal gravity, the finiteness of Euclidean on-shell action and conserved charges for solutions with AdS asymptotics have been shown explicitly in four~\cite{Grumiller:2013mxa} and in six dimensions~\cite{Anastasiou:2020mik,Anastasiou:2021tlv}. Nevertheless, a rigorous proof of the finiteness of the latter is still lacking; at least in the weakly modified asymptotically AdS sector. However, some advances have been made towards understanding how the curvature falloff restricts the sub-leading deformations of the metric with respect to the Einstein sector~\cite{Ghodsi:2014hua,Anastasiou:2016jix}. Indeed, Maldacena proposed a simple Neumann boundary condition that selects Einstein spaces as solutions of conformal gravity~\cite{Maldacena:2011mk} (see also~\cite{Anastasiou:2020mik}). In this section, we provide additional evidence for the relation between renormalization and conformal invariance, in this case, in presence of scalar fields. We shall use the Noether-Wald formalism~\cite{Wald:1993nt,Iyer:1994ys} to show that conserved charges are finite without any reference to intrinsic boundary counterterms; only the generalized Gibbons--Hawking--York will be kept. However, the latter vanishes for the instantonic solution of conformally invariant Brans-Dicke gravity presented in Eq.~\eqref{sol2}, in contrast to Eq.~\eqref{solnutconformal} whose contribution is nontrivial. 

Diffeomorphism invariance of the action principle generated by a vector field $\xi=\xi^\mu\partial_\mu$, implies the local identity $\nabla_\mu J^\mu=-\Lie_\xi g^{\mu\nu}\mathcal{E}_{\mu\nu}$, where $\tilde{\mathcal{E}}_{\mu\nu}=0$ are the equations of motion for the metric defined at the beginning of this Section, $\Lie_\xi$ is the Lie derivative along the vector field $\xi$, $J^\mu$ is the Noether current defined through
\begin{align}
    J^\mu = -2\nabla_\nu\left(E^{\mu\nu}_{\lambda\rho}\nabla^\lambda\xi^\rho + 2\xi^\lambda\nabla^{\rho}E^{\mu\nu}_{\lambda\rho} \right) \;\;\;\;\; \mbox{and} \;\;\;\;\; E^{\mu\nu}_{\lambda\rho} = -\frac{1}{24}\phi^2 \delta^{\mu\nu}_{\lambda\rho}\,.
\end{align}
When the field equations hold, the Noether current is conserved, i.e.,  $\nabla_\mu J^\mu=0$. The Poincaré Lemma, in turn, implies that the latter can be written locally as $J^\mu=\nabla_\nu q^{\mu\nu}$, which implicitly defines the Noether prepotential $q^{\mu\nu}=-q^{\nu\mu}$. When $\xi$ is a Killing vector, integrating the Noether prepotential over a codimension-2 hypersurface $\Sigma$ gives the Noether charge associated to $\xi$, namely,
\begin{align}
    Q[\xi] = \frac{1}{2}\int_\Sigma\epsilon_{\mu\nu\lambda\rho}q^{\mu\nu}\diff{x^\lambda}\wedge\diff{x^\rho} \equiv \int_\Sigma Q_{\mu\nu}\diff{x^\mu}\wedge\diff{x^\nu}\,.
\end{align}

Since the (anti-)self dual Eguchi-Hanson instanton is endowed with a certain degree of anisotropy, we postpone a deeper study of its conserved charges for the future. Here, we focus on the novel Taub-NUT instanton in Eq.~\eqref{solnutconformal} whose asymptotically locally AdS behavior is well known. Let us consider the Killing vector field associated to Euclidean time symmetry, namely, $\xi=\partial_\tau$. The relevant components of the dual Noether prepotential $2$-form are %for the Eguchi-Hanson metric in Eq.~\eqref{Eguchi-Hanson-ansatz} are
% \begin{align}
%     Q_{\vartheta\varphi}^{\rm (EH)}[\xi] &= \frac{r}{8}\left[\kappa\,\frac{\diff{}}{\diff{r}}\left(r^2f \right) - \frac{\phi^6}{12}\,\frac{\diff{}}{\diff{r}}\left(\frac{r^2f}{\phi^4}\right) \right]\sin\vartheta\,, \\  Q_{r\varphi}^{\rm (EH)}[\xi] &= -\frac{rf}{2}\left[\kappa-\frac{\phi^2}{12}\right]\cos\vartheta\,,
% \end{align}
%while for the Taub-NUT metric in Eq.~\eqref{Taub-NUT-ansatz} are given by
\begin{align}
    Q_{\vartheta\varphi} &= -\frac{\left(r^2-n^2\right)\phi^6}{12} \,\frac{\diff{}}{\diff{r}}\left(\frac{f}{\phi^4}\right) \sin\vartheta\,, \\
    Q_{r\varphi} &= \frac{n^2f\phi^2}{3\left(r^2-n^2\right)}\cos\vartheta\,.
\end{align}
The mass of the Taub-NUT instanton can be computed by means of the asymptotic charge associated to the Euclidean time symmetry generated by $\xi=\partial_\tau$, that is~\cite{Wald:1993nt,Iyer:1994ys}
\begin{align}
    M &= \int_\infty\left(Q_{\mu\nu} - \xi^\lambda B_{\lambda\mu\nu} \right)\diff{x^\mu}\wedge\diff{x^\nu} \,,
\end{align}
where $B_{\lambda\mu\nu}=-\frac{1}{6}\phi^2 K n^\rho\epsilon_{\lambda\mu\nu\rho}$ is the generalized GHY term that guarantees a well-posed variational principle (see for instance~\cite{Padilla:2012ze}). Focusing on the conformally invariant case, the mass of the Taub-NUT instanton in Eq.~\eqref{solnutconformal} at the (anti-)self-dual point, namely when Eq.~\eqref{bandphifornutconformal} holds, is given by 
% \begin{align}
%     %Q_{\rm EH}[\xi] &= 0\,, \\
%     \label{mass_NUT} 
%     M &= -\frac{2\pi\,\mu\,\lambda\left(\mu-n\right)\left[1+\frac{\lambda}{3}\left(\mu-n\right)^2 \right]}{9\alpha\left(\mu-3n\right)}\,,
% \end{align}
% for the Eguchi-Hanson and Taub-NUT instantons, respectively, in the conformally invariant case of the action~\eqref{Ibulk}, recall, $\kappa\to0$. 
% 
% In the Taub-NUT case, the asymptotic charge associated to the Euclidean time symmetry is the ADM mass~\cite{Garfinkle:2000ms,Ciambelli:2020qny}, that is, $Q_{\rm NUT}[\xi] = M_{\rm NUT}$. Thus, in the (anti-)self dual case, namely, when Eq.~\eqref{bandphifornutconformal} holds, the mass can be written in terms of $\phi_0$ as 
\begin{align}
    M_{\text{NUT}}=\frac{4\pi\mu\lambda\phi_{0}^2}{9}\,.
\end{align}
A desirable physical condition is that the mass of the Taub-NUT instanton is bounded from below. First, we observe that if $\lambda \geq 0$ or $\lambda<0$ the solution is either asymptotically locally dS or AdS, respectively. For $\lambda \geq 0$, the condition $\phi_0\in \mathbb{R}$ implies the following constraints
\begin{align}
    0<\mu<3n \qquad \text{and} \qquad 0<M_{\rm NUT}<+\infty\,.
\end{align}
Thus, it is clear that this condition guarantees a region in the parameter space where the mass is bounded from below. For $\lambda<0$, on the other hand, reality of the scalar field leads to four different cases, they are
% \begin{itemize}
%     \item $\mu>3n \; \wedge \; \lambda<-\frac{3}{(\mu-n)^2}\;$ implies $-\infty<M_{\rm NUT}<0$. %since $\phi_0^2 \to +\infty$ as $\mu \to 3n$.
%     \item $n<\mu<3n\; \wedge \; -\frac{3}{(\mu-n)^2}<\lambda<0\;$ implies $-\infty<M_{\rm NUT}<0$. %since $\phi_0^2 \to +\infty$ as $\mu \to 3n$.
%     \item $\mu<n \;\; \mbox{and} \;\; \lambda<-\frac{3}{(\mu-n)^2}$ yields $-\infty<M_{\rm NUT}<0$ since $\lambda \to -\infty$ as $\mu \to n$.
% \end{itemize}
\begin{align*}
    &\mbox{(a)}& \mu>&\,3n&  &\wedge&  \lambda&<-\frac{3}{(\mu-n)^2}& &\mbox{implies}& -\infty<M_{\rm NUT}<0\,, \\
    &\mbox{(b)}& n<\mu<&\,3n&  &\wedge&  -&\frac{3}{(\mu-n)^2}<\lambda<0& &\mbox{implies}& -\infty<M_{\rm NUT}<0\,, \\
    &\mbox{(c)}& 0<\mu<&\,n&  &\wedge&  \lambda&<-\frac{3}{(\mu-n)^2}& &\mbox{implies}& -\infty<M_{\rm NUT}<0\,,\\
    &\mbox{(d)}& \mu<0 \quad\wedge\quad &\mu<\,n&  &\wedge&  \lambda&<-\frac{3}{(\mu-n)^2}& &\mbox{implies}& 0<M_{\rm NUT}<+\infty\,.
\end{align*}
Therefore, we conclude that the case (d) yields a mass which is bounded from below while possessing an asymptotically locally AdS behavior and a real scalar field.

The entropy, on the other hand, is obtained by integrating the Noether charge over the codimension-2 hypersurfaces that represent obstructions to the foliation with a time function that provides the unitary Hamiltonian evolution of the system~\cite{Hawking:1998jf}. For black holes, this is given by the event horizon. For the Taub-NUT instanton, however, there appear additional obstructions coming from the Misner string~\cite{Hawking:1998jf,Garfinkle:2000ms,Astefanesei:2004ji,Hawking:1998ct,Ciambelli:2020qny,Emparan:1999pm,Mann:1999pc}. Thus, using the generalized entropy formula presented in Ref.~\cite{Garfinkle:2000ms} (see also~\cite{Ciambelli:2020qny}), namely,
\begin{align}
    S = \beta_\tau\int_0^{2\pi}\diff{\phi}\left[\int_0^{\pi}\diff{\vartheta}\,Q_{\vartheta\varphi}\Big|_{r=r_b} + \int_{r_b}^\infty\diff{r}\,Q_{r\varphi}\Big|_{\vartheta=\pi} - \int_{r_b}^\infty\diff{r}\,Q_{r\varphi}\Big|_{\vartheta=0} \right]\,,
\end{align}
we find that the entropy of the (anti-)self dual 
Taub-NUT solution of Eq.~\eqref{solnutconformal} is 
\begin{align}
    %S_{\rm EH} &= \frac{\pi^2}{6\alpha}\left(2-\frac{b}{r_b^2} \right)\,, \\
    %S_{\rm NUT} &= \frac{8\pi^2 n^2\left[1+\frac{2\lambda n}{3}\left(\mu-n\right)\right]\left[1+\frac{\lambda}{3}\left(\mu-n\right)^2 \right]}{3\alpha\left(\mu-3n\right)^2}\,, \\
    S_{\rm NUT} &= -\frac{16\pi^2 n^2\phi_0^2\left[3+2\lambda n\left(\mu-n\right)\right]}{9(\mu-n)\left(\mu-3n\right)}\,, %\\
    %\frac{2\phi_0^2}{3(\mu-n)\left(\mu-3n\right)}  &= -\frac{\left[3+\lambda\left(\mu-n\right)^2\right]}{9\alpha\left(\mu-3n\right)^2}\,.
\end{align}
where the conditions~\eqref{bandphifornutconformal} have been used. Since the boundedness of the mass for the asymptotically locally AdS Taub-NUT instanton restricts the parameter space accordingly to the case (d), it is direct to see that the entropy is strictly positive.

%\occ{Marcelo:(I don't know if we'll say sommething about the thermodynamics of the Einstein sector) Profs, I computed the Wald entropy for the solution \eqref{sol1a} considering the contributions coming from the Misner strings. It is finite and given by $S_{EH_{\kappa\neq0}}=\frac{\pi r_b^2}{4G}-\frac{\pi^2b}{3\alpha r_b^2}$. It is almost the law of a quarter of the horizon area.  If we will add it, can you check that?  Thank you!   }

The Pontryagin density with fixed coupling in the action~\eqref{Ibulk} allows us to compute the conserved charges associated to the Maxwell sector through the Noether formalism developed in Ref.~\cite{Araneda:2016iiy}. Indeed, the particular choice of the $\Theta$-parameter yields
\begin{align}
    \mathcal{Q}[\lambda] = -\int \left(F^{\mu\nu}\pm \tilde{F}^{\mu\nu} \right)\lambda\,\diff{\Sigma_{\mu\nu}}\,,
\end{align}
where $\diff{\Sigma_{\mu\nu}}$ is the area element of the codimension-2 hypersurface. Notice that the $U(1)$ conserved charge vanishes identically at the (anti-)self-dual point. Asymptotically, the gauge parameter $\lambda$ can be normalized to $1$ without loss of generality. Then, the conserved charge associated to the Maxwell field supporting the Taub-NUT solution in Eq.~\eqref{solnutconformal} is 
\begin{align}
    %\mathcal{Q}_{\rm EH} = 0\;\;\;\;\; \mbox{and} \;\;\;\;\;  
    \mathcal{Q}_{\rm NUT} = 4\pi\left(p-q\right)\,.
\end{align}
From this expression it is clear to see that the (anti-)self-dual point of the Maxwell field is achieved when $p=q$. In the Eguchi-Hanson case, on the other hand, the Maxwell field is globally (anti-)self dual, which implies that $\mathcal{Q}_{\rm EH}=0$. Indeed, the Chern-Pontryagin index of $U(1)$ in this case [see Eq.~\eqref{sol2}] is
\begin{align}
   c=-\frac{1}{8\pi^2}\int_{\mathcal{M}}\diff{^4}x\,\sqrt{|g|}\;\tilde{F}_{\mu\nu}F^{\mu\nu} =  \frac{4\,q^2}{r_b^2\left(2r_b^2+b\right)}\,.
\end{align}
Thus, we conclude that the contribution of Maxwell fields to the Euclidean on-shell action in Eq.~\eqref{IEHconformal} is purely topological over an Eguchi-Hanson space. This can be seen because, even though their Noether charge vanishes identically due to (anti-)self duality, their topological charge is nontrivial.

\section{Discussion~\label{sec:conclusions}}

In this work, we study different charged gravitational instantons in general relativity with a conformally coupled scalar field. In particular, we focus on the four-dimensional Eguchi-Hanson and Taub-NUT Euclidean spaces build upon the $U(1)$ fibration of two-dimensional Einstein-K\"ahler manifolds. Different regular solutions of this kind are analyzed, representing topologically nontrivial configurations labeled by the Hirzebruch signature and Euler characteristic. Additionally, we obtain the period of the Euclidean time such that the geometry is geodesically complete. The inverse of the latter is interpreted as the Hawking temperature of these instantons. We also uncover additional interesting properties that we summarize next.

First, we find that the two-parameter family of Maxwell fields is not (anti-)self dual in the Eguchi-Hanson space. Indeed, the general solution of the gauge potential introduces a divergent term at the action level and variations thereof that can be cured by adding the $U(1)$ Chern-Pontryagin density with a particular coupling. The latter renders the electromagnetic ground state as a globally (anti-)self-dual configuration (see Ref.~\cite{Miskovic:2009bm,Araneda:2016iiy} for details) which is included as a particular case of the general solution found in Eq.~\eqref{sol1}. In the Taub-NUT case, however, the (anti-)self-dual point trivializes the scalar field and the Maxwell field becomes (anti-)self dual if and only if the electric and magnetic charge are equal. In that case, one recovers the Brill solution~\cite{PhysRev.133.B845} at the (anti-)self-dual point. In the Taub-Bolt case, on the other hand, the solution is endowed with a nontrivial scalar and Maxwell fields that lead to an asymptotically locally (anti-)self-dual space. 

In presence of higher-curvature corrections, we obtain a novel solution of the Eguchi-Hanson type with a nontrivial scalar and Maxwell fields. In this case, the new coupling constants open a window where the Maxwell field does not introduce divergent pieces at the action level. Indeed, we obtain a solution with a negative cosmological constant and a positive curvature K\"ahler base manifold due to the higher-curvature corrections; something that it is not guaranteed in their absence.

The conformally invariant case, on the other hand, leads to a particular case of Brans-dicke gravity where the Einstein sector decouples from the scalar-tensor one. We show explicitly that the Euclidean on-shell action of the two novel gravitational instantons in Sec.~\ref{sec:conformal} are finite. Moreover, we prove that the globally (anti-)self-dual Taub-NUT metric with nontrivial conformally coupled scalar field possesses finite conserved charges as well. This provides additional evidence of the relation between finiteness and conformal invariance for asymptotically locally AdS spaces in presence of scalar and Maxwell fields.

Finally, we reveal interesting avenues that are worth exploring in future works. First, the presence of a negative cosmological constant implies that the Eguchi-Hanson metric is not asymptotically conformally flat. The solution, however, is completely regular as far as the absence of conical singularities at the bolt is assumed. In presence of the Einstein term, the computation of its Euclidean on-shell action and conserved charges will certainly provide valuable information in order to understand its thermodynamics and phase transitions. The Taub-NUT space, in contrast, is indeed asymptotically locally AdS. Therefore, its thermodynamics can be worked out using standard Euclidean techniques at first-order in the saddle point approximation. We postpone a deeper study of of these subjects for the future.   
\vspace*{0.3cm}

\emph{Note added: A misprint has been corrected in Eqs.~\eqref{Diracindex} and~\eqref{indexnutconf} with respect to the published version.}

\acknowledgments
%\begin{acknowledgments}
We thank G.~Anastasiou, I.~J.~Araya, D.~Flores-Alfonso, G.~Giribet, N.~Mora, R.~Olea, and the anonymous referee for insightful discussions, valuable comments and remarks. The work of J. B. is supported by the  ``Programme to support prospective human resources -- post Ph.D. candidates''  of the Czech Academy of Sciences, project L100192101. A. C. and C. C. are partially supported by Agencia Nacional de Investigación y Desarrollo (ANID) through FONDECYT grants No~1210500 and~11200025, respectively. M.~O. is supported by Beca ANID de Mag\'{\i}ster 22201618.
%\end{acknowledgments}

%\pagebreak

\bibliography{References}

\providecommand{\href}[2]{#2}\begingroup\raggedright\begin{thebibliography}{100}

\bibitem{Hawking:1976jb}
S.~W. Hawking,\,
  \href{http://dx.doi.org/10.1016/0375-9601(77)90386-3}{\emph{Phys. Lett. A}
  {\bfseries 60} (1977) 81}.

\bibitem{Gibbons:1978tef}
G.~W. Gibbons and S.~W. Hawking,\,
  \href{http://dx.doi.org/10.1016/0370-2693(78)90478-1}{\emph{Phys. Lett. B}
  {\bfseries 78} (1978) 430}.

\bibitem{Hawking:1978ghb}
S.~W. Hawking and C.~N. Pope,\,
  \href{http://dx.doi.org/10.1016/0550-3213(78)90073-1}{\emph{Nucl. Phys. B}
  {\bfseries 146} (1978) 381--392}.

\bibitem{Eguchi:1978xp}
T.~Eguchi and A.~J. Hanson,\,
  \href{http://dx.doi.org/10.1016/0370-2693(78)90566-X}{\emph{Phys. Lett. B}
  {\bfseries 74} (1978) 249--251}.

\bibitem{Eguchi:1978gw}
T.~Eguchi and A.~J. Hanson,\,
  \href{http://dx.doi.org/10.1016/0003-4916(79)90282-3}{\emph{Annals Phys.}
  {\bfseries 120} (1979) 82}.

\bibitem{Eguchi:1980jx}
T.~Eguchi, P.~B. Gilkey and A.~J. Hanson,\,
  \href{http://dx.doi.org/10.1016/0370-1573(80)90130-1}{\emph{Phys. Rept.}
  {\bfseries 66} (1980) 213}.

\bibitem{Alvarez-Gaume:1983ihn}
L.~Alvarez-Gaume and E.~Witten,\,
  \href{http://dx.doi.org/10.1016/0550-3213(84)90066-X}{\emph{Nucl. Phys. B}
  {\bfseries 234} (1984) 269}.

\bibitem{Witten:1985xe}
E.~Witten,\, \href{http://dx.doi.org/10.1007/BF01212448}{\emph{Commun. Math.
  Phys.} {\bfseries 100} (1985) 197}.

\bibitem{Linshaw:2017bpf}
A.~Linshaw and V.~Mathai,\,
  \href{http://dx.doi.org/10.1016/j.geomphys.2018.03.017}{\emph{J. Geom. Phys.}
  {\bfseries 129} (2018) 269--278}.

\bibitem{Hashemi:2018jbv}
S.~S. Hashemi and N.~Riazi,\,
  \href{http://dx.doi.org/10.1016/j.aop.2018.04.004}{\emph{Annals Phys.}
  {\bfseries 393} (2018) 206--214}.

\bibitem{Li2019}
Y.~Li,\, \href{http://dx.doi.org/10.1007/s00222-019-00861-w}{\emph{Inventiones
  mathematicae} {\bfseries 217} (Jan, 2019) 1}.

\bibitem{Taub:1950ez}
A.~H. Taub,\, \href{http://dx.doi.org/10.2307/1969567}{\emph{Annals Math.}
  {\bfseries 53} (1951) 472--490}.

\bibitem{Newman:1963yy}
E.~Newman, L.~Tamburino and T.~Unti,\,
  \href{http://dx.doi.org/10.1063/1.1704018}{\emph{J. Math. Phys.} {\bfseries
  4} (1963) 915}.

\bibitem{Page:1978hdy}
D.~N. Page,\, \href{http://dx.doi.org/10.1016/0370-2693(78)90016-3}{\emph{Phys.
  Lett. B} {\bfseries 78} (1978) 249--251}.

\bibitem{Gibbons:1979xm}
G.~W. Gibbons and S.~W. Hawking,\,
  \href{http://dx.doi.org/10.1007/BF01197189}{\emph{Commun. Math. Phys.}
  {\bfseries 66} (1979) 291--310}.

\bibitem{Lynden-Bell:1996dpw}
D.~Lynden-Bell and M.~Nouri-Zonoz,\,
  \href{http://dx.doi.org/10.1103/RevModPhys.70.427}{\emph{Rev. Mod. Phys.}
  {\bfseries 70} (1998) 427--446}.

\bibitem{Bicak:2000ea}
J.~Bicak,\, {\emph{Lect. Notes Phys.} {\bfseries 540} (2000) 1--126}.

\bibitem{Miskovic:2009bm}
O.~Miskovic and R.~Olea,\,
  \href{http://dx.doi.org/10.1103/PhysRevD.79.124020}{\emph{Phys. Rev. D}
  {\bfseries 79} (2009) 124020}.

\bibitem{Araneda:2016iiy}
R.~Araneda, R.~Aros, O.~Miskovic and R.~Olea,\,
  \href{http://dx.doi.org/10.1103/PhysRevD.93.084022}{\emph{Phys. Rev. D}
  {\bfseries 93} (2016) 084022}.

\bibitem{PhysRev.133.B845}
D.~R. Brill,\, \href{http://dx.doi.org/10.1103/PhysRev.133.B845}{\emph{Phys.
  Rev.} {\bfseries 133} (Feb, 1964) B845--B848}.

\bibitem{Mann:2020wad}
R.~B. Mann, L.~A. Pando~Zayas and M.~Park,\,
  \href{http://dx.doi.org/10.1007/JHEP03(2021)039}{\emph{JHEP} {\bfseries 03}
  (2021) 039}.

\bibitem{Abbasvandi:2021nyv}
N.~Abbasvandi, M.~Tavakoli and R.~B. Mann,\,
  \href{http://dx.doi.org/10.1007/JHEP08(2021)152}{\emph{JHEP} {\bfseries 08}
  (2021) 152}.

\bibitem{Bandos:2020jsw}
I.~Bandos, K.~Lechner, D.~Sorokin and P.~K. Townsend,\,
  \href{http://dx.doi.org/10.1103/PhysRevD.102.121703}{\emph{Phys. Rev. D}
  {\bfseries 102} (2020) 121703}.

\bibitem{BallonBordo:2020jtw}
A.~Ballon~Bordo, D.~Kubiz\v{n}\'ak and T.~R. Perche,\,
  \href{http://dx.doi.org/10.1016/j.physletb.2021.136312}{\emph{Phys. Lett. B}
  {\bfseries 817} (2021) 136312}.

\bibitem{Flores-Alfonso:2020nnd}
D.~Flores-Alfonso, R.~Linares and M.~Maceda,\,
  \href{http://dx.doi.org/10.1007/JHEP09(2021)104}{\emph{JHEP} {\bfseries 09}
  (2021) 104}.

\bibitem{Zhang:2021qga}
M.~Zhang and J.~Jiang,\,
  \href{http://dx.doi.org/10.1103/PhysRevD.104.084094}{\emph{Phys. Rev. D}
  {\bfseries 104} (2021) 084094}.

\bibitem{Sakti:2019udk}
M.~F. Sakti, A.~Suroso and F.~P. Zen,\,
  \href{http://dx.doi.org/10.1140/epjp/i2019-12937-x}{\emph{Eur. Phys. J. Plus}
  {\bfseries 134} (2019) 580}.

\bibitem{Sakti:2019zix}
M.~F. Sakti, A.~M. Ghezelbash, A.~Suroso and F.~P. Zen,\,
  \href{http://dx.doi.org/10.1007/s10714-019-2641-z}{\emph{Gen. Rel. Grav.}
  {\bfseries 51} (2019) 151}.

\bibitem{Sakti:2019krw}
M.~F. Sakti, A.~Suroso and F.~P. Zen,\,
  \href{http://dx.doi.org/10.1016/j.aop.2019.168062}{\emph{Annals Phys.}
  {\bfseries 413} (2020) 168062}.

\bibitem{Sakti:2020jpo}
M.~F. Sakti, A.~M. Ghezelbash, A.~Suroso and F.~P. Zen,\,
  \href{http://dx.doi.org/10.1016/j.nuclphysb.2020.114970}{\emph{Nucl. Phys. B}
  {\bfseries 953} (2020) 114970}.

\bibitem{Leigh:2011au}
R.~G. Leigh, A.~C. Petkou and P.~M. Petropoulos,\,
  \href{http://dx.doi.org/10.1103/PhysRevD.85.086010}{\emph{Phys. Rev. D}
  {\bfseries 85} (2012) 086010}.

\bibitem{Leigh:2012jv}
R.~G. Leigh, A.~C. Petkou and P.~M. Petropoulos,\,
  \href{http://dx.doi.org/10.1007/JHEP11(2012)121}{\emph{JHEP} {\bfseries 11}
  (2012) 121}.

\bibitem{Caldarelli:2012cm}
M.~M. Caldarelli, R.~G. Leigh, A.~C. Petkou, P.~M. Petropoulos, V.~Pozzoli and
  K.~Siampos,\, \href{http://dx.doi.org/10.22323/1.155.0076}{\emph{PoS}
  {\bfseries CORFU2011} (2011) 076}.

\bibitem{Mukhopadhyay:2013gja}
A.~Mukhopadhyay, A.~C. Petkou, P.~M. Petropoulos, V.~Pozzoli and K.~Siampos,\,
  \href{http://dx.doi.org/10.1007/JHEP04(2014)136}{\emph{JHEP} {\bfseries 04}
  (2014) 136}.

\bibitem{Kalamakis:2020aaj}
G.~Kalamakis, R.~G. Leigh and A.~C. Petkou,\,
  \href{http://dx.doi.org/10.1103/PhysRevD.103.126012}{\emph{Phys. Rev. D}
  {\bfseries 103} (2021) 126012}.

\bibitem{Misner:1963fr}
C.~W. Misner,\, \href{http://dx.doi.org/10.1063/1.1704019}{\emph{J. Math.
  Phys.} {\bfseries 4} (1963) 924--938}.

\bibitem{Hennigar:2019ive}
R.~A. Hennigar, D.~Kubiz\v{n}\'ak and R.~B. Mann,\,
  \href{http://dx.doi.org/10.1103/PhysRevD.100.064055}{\emph{Phys. Rev. D}
  {\bfseries 100} (2019) 064055}.

\bibitem{Chen:2019uhp}
Z.~Chen and J.~Jiang,\,
  \href{http://dx.doi.org/10.1103/PhysRevD.100.104016}{\emph{Phys. Rev. D}
  {\bfseries 100} (2019) 104016}.

\bibitem{BallonBordo:2020mcs}
A.~Ballon~Bordo, F.~Gray and D.~Kubiz\v{n}\'ak,\,
  \href{http://dx.doi.org/10.1007/JHEP05(2020)084}{\emph{JHEP} {\bfseries 05}
  (2020) 084}.

\bibitem{Cano:2021qzp}
P.~A. Cano and D.~Pere\~niguez,\,
  \href{https://arxiv.org/abs/2101.10652}{{\ttfamily 2101.10652}}.

\bibitem{Clement:2015cxa}
G.~Cl\'ement, D.~Gal'tsov and M.~Guenouche,\,
  \href{http://dx.doi.org/10.1016/j.physletb.2015.09.074}{\emph{Phys. Lett. B}
  {\bfseries 750} (2015) 591--594}.

\bibitem{Clement:2015aka}
G.~Cl\'ement, D.~Gal'tsov and M.~Guenouche,\,
  \href{http://dx.doi.org/10.1103/PhysRevD.93.024048}{\emph{Phys. Rev. D}
  {\bfseries 93} (2016) 024048}.

\bibitem{Hawking:1998jf}
S.~W. Hawking and C.~J. Hunter,\,
  \href{http://dx.doi.org/10.1103/PhysRevD.59.044025}{\emph{Phys. Rev. D}
  {\bfseries 59} (1999) 044025}.

\bibitem{Garfinkle:2000ms}
D.~Garfinkle and R.~B. Mann,\,
  \href{http://dx.doi.org/10.1088/0264-9381/17/16/314}{\emph{Class. Quant.
  Grav.} {\bfseries 17} (2000) 3317--3324}.

\bibitem{Astefanesei:2004ji}
D.~Astefanesei, R.~B. Mann and E.~Radu,\,
  \href{http://dx.doi.org/10.1016/j.physletb.2005.05.057}{\emph{Phys. Lett. B}
  {\bfseries 620} (2005) 1--8}.

\bibitem{Hawking:1998ct}
S.~W. Hawking, C.~J. Hunter and D.~N. Page,\,
  \href{http://dx.doi.org/10.1103/PhysRevD.59.044033}{\emph{Phys. Rev. D}
  {\bfseries 59} (1999) 044033}.

\bibitem{Emparan:1999pm}
R.~Emparan, C.~V. Johnson and R.~C. Myers,\,
  \href{http://dx.doi.org/10.1103/PhysRevD.60.104001}{\emph{Phys. Rev. D}
  {\bfseries 60} (1999) 104001}.

\bibitem{Mann:1999pc}
R.~B. Mann,\, \href{http://dx.doi.org/10.1103/PhysRevD.60.104047}{\emph{Phys.
  Rev. D} {\bfseries 60} (1999) 104047}.

\bibitem{Ciambelli:2020qny}
L.~Ciambelli, C.~Corral, J.~Figueroa, G.~Giribet and R.~Olea,\,
  \href{http://dx.doi.org/10.1103/PhysRevD.103.024052}{\emph{Phys. Rev. D}
  {\bfseries 103} (2021) 024052}.

\bibitem{Johnson:2014xza}
C.~V. Johnson,\,
  \href{http://dx.doi.org/10.1088/0264-9381/31/23/235003}{\emph{Class. Quant.
  Grav.} {\bfseries 31} (2014) 235003}.

\bibitem{Johnson:2014pwa}
C.~V. Johnson,\,
  \href{http://dx.doi.org/10.1088/0264-9381/31/22/225005}{\emph{Class. Quant.
  Grav.} {\bfseries 31} (2014) 225005}.

\bibitem{Belavin:1975fg}
A.~A. Belavin, A.~M. Polyakov, A.~S. Schwartz and Y.~S. Tyupkin,\,
  \href{http://dx.doi.org/10.1016/0370-2693(75)90163-X}{\emph{Phys. Lett. B}
  {\bfseries 59} (1975) 85--87}.

\bibitem{Xiao_2004}
Z.~Xiao,\,
  \href{http://dx.doi.org/10.1088/0253-6102/42/2/235}{\emph{Communications in
  Theoretical Physics} {\bfseries 42} (aug, 2004) 235--238}.

\bibitem{Chen:2020org}
J.~Chen and X.~Zhang,\,
  \href{http://dx.doi.org/10.1016/j.geomphys.2020.104010}{\emph{J. Geom. Phys.}
  {\bfseries 161} (2021) 104010}.

\bibitem{Corral:2021xsu}
C.~Corral, G.~Giribet and R.~Olea,\,
  \href{http://dx.doi.org/10.1103/PhysRevD.104.064026}{\emph{Phys. Rev. D}
  {\bfseries 104} (2021) 064026}.

\bibitem{Burgess:1994kq}
C.~P. Burgess, R.~C. Myers and F.~Quevedo,\,
  \href{http://dx.doi.org/10.1016/S0550-3213(95)00090-9}{\emph{Nucl. Phys.}
  {\bfseries B442} (1995) 75--96}.

\bibitem{Johnson:1994ek}
C.~V. Johnson and R.~C. Myers,\,
  \href{http://dx.doi.org/10.1103/PhysRevD.50.6512}{\emph{Phys. Rev.}
  {\bfseries D50} (1994) 6512--6518}.

\bibitem{Johnson:1994nj}
C.~V. Johnson and R.~C. Myers,\, , \emph{{Stringy twists of the Taub - NUT
  metric}},  in \emph{{7th Marcel Grossmann Meeting on General Relativity (MG
  7)}}, pp.~940--942, 7, 1994.
\newblock \href{https://arxiv.org/abs/hep-th/9409177}{{\ttfamily
  hep-th/9409177}}.

\bibitem{Strominger:1984zy}
A.~Strominger, G.~T. Horowitz and M.~J. Perry,\,
  \href{http://dx.doi.org/10.1016/0550-3213(84)90340-7}{\emph{Nucl. Phys. B}
  {\bfseries 238} (1984) 653--664}.

\bibitem{Dehghani:2005zm}
M.~H. Dehghani and R.~B. Mann,\,
  \href{http://dx.doi.org/10.1103/PhysRevD.72.124006}{\emph{Phys. Rev.}
  {\bfseries D72} (2005) 124006}.

\bibitem{Dehghani:2006aa}
M.~H. Dehghani and S.~H. Hendi,\,
  \href{http://dx.doi.org/10.1103/PhysRevD.73.084021}{\emph{Phys. Rev.}
  {\bfseries D73} (2006) 084021}.

\bibitem{Hendi:2008wq}
S.~H. Hendi and M.~H. Dehghani,\,
  \href{http://dx.doi.org/10.1016/j.physletb.2008.07.002}{\emph{Phys. Lett.}
  {\bfseries B666} (2008) 116--120}.

\bibitem{Bueno:2018uoy}
P.~Bueno, P.~A. Cano, R.~A. Hennigar and R.~B. Mann,\,
  \href{http://dx.doi.org/10.1007/JHEP10(2018)095}{\emph{JHEP} {\bfseries 10}
  (2018) 095}.

\bibitem{Corral:2019leh}
C.~Corral, D.~Flores-Alfonso and H.~Quevedo,\,
  \href{http://dx.doi.org/10.1103/PhysRevD.100.064051}{\emph{Phys. Rev.}
  {\bfseries D100} (2019) 064051}.

\bibitem{Grumiller:2013mxa}
D.~Grumiller, M.~Irakleidou, I.~Lovrekovic and R.~McNees,\,
  \href{http://dx.doi.org/10.1103/PhysRevLett.112.111102}{\emph{Phys. Rev.
  Lett.} {\bfseries 112} (2014) 111102}.

\bibitem{Carter:1971zc}
B.~Carter,\, \href{http://dx.doi.org/10.1103/PhysRevLett.26.331}{\emph{Phys.
  Rev. Lett.} {\bfseries 26} (1971) 331--333}.

\bibitem{Israel:1967wq}
W.~Israel,\, \href{http://dx.doi.org/10.1103/PhysRev.164.1776}{\emph{Phys.
  Rev.} {\bfseries 164} (1967) 1776--1779}.

\bibitem{Israel:1967za}
W.~Israel,\, \href{http://dx.doi.org/10.1007/BF01645859}{\emph{Commun. Math.
  Phys.} {\bfseries 8} (1968) 245--260}.

\bibitem{Wald:1971iw}
R.~M. Wald,\, \href{http://dx.doi.org/10.1103/PhysRevLett.26.1653}{\emph{Phys.
  Rev. Lett.} {\bfseries 26} (1971) 1653--1655}.

\bibitem{Herdeiro:2015waa}
C.~A.~R. Herdeiro and E.~Radu,\,
  \href{http://dx.doi.org/10.1142/S0218271815420146}{\emph{Int. J. Mod. Phys.
  D} {\bfseries 24} (2015) 1542014}.

\bibitem{Bekenstein:1974sf}
J.~D. Bekenstein,\,
  \href{http://dx.doi.org/10.1016/0003-4916(74)90124-9}{\emph{Annals Phys.}
  {\bfseries 82} (1974) 535--547}.

\bibitem{Bocharova:1970skc}
N.~M. Bocharova, K.~A. Bronnikov and V.~N. Melnikov,\, {\emph{Vestn. Mosk.
  Univ. Fiz. Astron.} {\bfseries 6} (1970) 706}.

\bibitem{Martinez:2002ru}
C.~Martinez, R.~Troncoso and J.~Zanelli,\,
  \href{http://dx.doi.org/10.1103/PhysRevD.67.024008}{\emph{Phys. Rev. D}
  {\bfseries 67} (2003) 024008}.

\bibitem{Martinez:2005di}
C.~Martinez, J.~P. Staforelli and R.~Troncoso,\,
  \href{http://dx.doi.org/10.1103/PhysRevD.74.044028}{\emph{Phys. Rev. D}
  {\bfseries 74} (2006) 044028}.

\bibitem{Charmousis:2009cm}
C.~Charmousis, T.~Kolyvaris and E.~Papantonopoulos,\,
  \href{http://dx.doi.org/10.1088/0264-9381/26/17/175012}{\emph{Class. Quant.
  Grav.} {\bfseries 26} (2009) 175012}.

\bibitem{Astorino:2013xxa}
M.~Astorino,\, \href{http://dx.doi.org/10.1103/PhysRevD.89.044022}{\emph{Phys.
  Rev. D} {\bfseries 89} (2014) 044022}.

\bibitem{Bardoux:2013swa}
Y.~Bardoux, M.~M. Caldarelli and C.~Charmousis,\,
  \href{http://dx.doi.org/10.1007/JHEP05(2014)039}{\emph{JHEP} {\bfseries 05}
  (2014) 039}.

\bibitem{Astorino:2014mda}
M.~Astorino,\, \href{http://dx.doi.org/10.1103/PhysRevD.91.064066}{\emph{Phys.
  Rev. D} {\bfseries 91} (2015) 064066}.

\bibitem{Anabalon:2009qt}
A.~Anabalon and H.~Maeda,\,
  \href{http://dx.doi.org/10.1103/PhysRevD.81.041501}{\emph{Phys. Rev. D}
  {\bfseries 81} (2010) 041501}.

\bibitem{Cisterna:2021xxq}
A.~Cisterna, A.~Neira-Gallegos, J.~Oliva and S.~C. Rebolledo-Caceres,\,
  \href{http://dx.doi.org/10.1103/PhysRevD.103.064050}{\emph{Phys. Rev. D}
  {\bfseries 103} (2021) 064050}.

\bibitem{Astorino:2013xc}
M.~Astorino,\, \href{http://dx.doi.org/10.1103/PhysRevD.87.084029}{\emph{Phys.
  Rev. D} {\bfseries 87} (2013) 084029}.

\bibitem{Astorino:2013sfa}
M.~Astorino,\, \href{http://dx.doi.org/10.1103/PhysRevD.88.104027}{\emph{Phys.
  Rev. D} {\bfseries 88} (2013) 104027}.

\bibitem{Caceres:2020myr}
N.~Caceres, J.~Figueroa, J.~Oliva, M.~Oyarzo and R.~Stuardo,\,
  \href{http://dx.doi.org/10.1007/JHEP04(2020)157}{\emph{JHEP} {\bfseries 04}
  (2020) 157}.

\bibitem{Anabalon:2012tu}
A.~Anabalon and A.~Cisterna,\,
  \href{http://dx.doi.org/10.1103/PhysRevD.85.084035}{\emph{Phys. Rev. D}
  {\bfseries 85} (2012) 084035}.

\bibitem{Ayon-Beato:2015ada}
E.~Ay\'on-Beato, M.~Hassa\"\i{}ne and J.~A. M\'endez-Zavaleta,\,
  \href{http://dx.doi.org/10.1103/PhysRevD.92.024048}{\emph{Phys. Rev. D}
  {\bfseries 92} (2015) 024048}.

\bibitem{Barrientos:2022avi}
J.~Barrientos, A.~Cisterna, N.~Mora and A.~Vigan\`o,\,
  \href{https://arxiv.org/abs/2202.06706}{{\ttfamily 2202.06706}}.

\bibitem{Brihaye:2016lsx}
Y.~Brihaye and E.~Radu,\,
  \href{http://dx.doi.org/10.1016/j.physletb.2016.11.055}{\emph{Phys. Lett. B}
  {\bfseries 764} (2017) 300--305}.

\bibitem{Arratia:2020hoy}
E.~Arratia, C.~Corral, J.~Figueroa and L.~Sanhueza,\,
  \href{http://dx.doi.org/10.1103/PhysRevD.103.064068}{\emph{Phys. Rev. D}
  {\bfseries 103} (2021) 064068}.

\bibitem{Bhattacharya:2013hvm}
S.~Bhattacharya and H.~Maeda,\,
  \href{http://dx.doi.org/10.1103/PhysRevD.89.087501}{\emph{Phys. Rev. D}
  {\bfseries 89} (2014) 087501}.

\bibitem{Eguchi:1976db}
T.~Eguchi and P.~G.~O. Freund,\,
  \href{http://dx.doi.org/10.1103/PhysRevLett.37.1251}{\emph{Phys. Rev. Lett.}
  {\bfseries 37} (1976) 1251}.

\bibitem{deHaro:2006ymc}
S.~de~Haro, I.~Papadimitriou and A.~C. Petkou,\,
  \href{http://dx.doi.org/10.1103/PhysRevLett.98.231601}{\emph{Phys. Rev.
  Lett.} {\bfseries 98} (2007) 231601}.

\bibitem{Anastasiou:2020mik}
G.~Anastasiou, I.~J. Araya and R.~Olea,\,
  \href{http://dx.doi.org/10.1007/JHEP01(2021)134}{\emph{JHEP} {\bfseries 01}
  (2021) 134}.

\bibitem{Anastasiou:2021tlv}
G.~Anastasiou, I.~J. Araya, C.~Corral and R.~Olea,\,
  \href{http://dx.doi.org/10.1007/JHEP07(2021)156}{\emph{JHEP} {\bfseries 2021}
  (2021) 156}.

\bibitem{Peccei:2006as}
R.~D. Peccei,\,
  \href{http://dx.doi.org/10.1007/978-3-540-73518-2_1}{\emph{Lect. Notes Phys.}
  {\bfseries 741} (2008) 3--17}.

\bibitem{Hartnoll:2005yc}
S.~A. Hartnoll and S.~P. Kumar,\,
  \href{http://dx.doi.org/10.1088/1126-6708/2005/06/012}{\emph{JHEP} {\bfseries
  06} (2005) 012}.

\bibitem{Bobev:2016sap}
N.~Bobev, T.~Hertog and Y.~Vreys,\,
  \href{http://dx.doi.org/10.1007/JHEP11(2016)140}{\emph{JHEP} {\bfseries 11}
  (2016) 140}.

\bibitem{Bueno:2018yzo}
P.~Bueno, P.~A. Cano, R.~A. Hennigar and R.~B. Mann,\,
  \href{http://dx.doi.org/10.1103/PhysRevLett.122.071602}{\emph{Phys. Rev.
  Lett.} {\bfseries 122} (2019) 071602}.

\bibitem{Pope:1981jx}
C.~N. Pope,\, \href{http://dx.doi.org/10.1088/0305-4470/14/5/007}{\emph{J.
  Phys. A} {\bfseries 14} (1981) L133--L137}.

\bibitem{Boulton:2021wln}
L.~Boulton, B.~Schroers and K.~Smedley-Williams,\,
  \href{https://arxiv.org/abs/2112.11411}{{\ttfamily 2112.11411}}.

\bibitem{APS-eta}
M.~F. Atiyah, V.~K. Patodi and I.~M. Singer,\,
  \href{http://dx.doi.org/https://doi.org/10.1112/blms/5.2.229}{\emph{Bulletin
  of the London Mathematical Society} {\bfseries 5} (1973) 229--234}.

\bibitem{Atiyah:2011fn}
M.~Atiyah, N.~S. Manton and B.~J. Schroers,\,
  \href{http://dx.doi.org/10.1098/rspa.2011.0616}{\emph{Proc. Roy. Soc. Lond.
  A} {\bfseries 468} (2012) 1252--1279}.

\bibitem{atiyah_lebrun_2013}
M.~Atiyah and C.~LeBrun,\,
  \href{http://dx.doi.org/10.1017/S0305004113000169}{\emph{Mathematical
  Proceedings of the Cambridge Philosophical Society} {\bfseries 155} (2013)
  13–37}.

\bibitem{Eguchi:1977iu}
T.~Eguchi, P.~B. Gilkey and A.~J. Hanson,\,
  \href{http://dx.doi.org/10.1103/PhysRevD.17.423}{\emph{Phys. Rev. D}
  {\bfseries 17} (1978) 423--427}.

\bibitem{Pope:1978zx}
C.~N. Pope,\, \href{http://dx.doi.org/10.1016/0550-3213(78)90038-X}{\emph{Nucl.
  Phys. B} {\bfseries 141} (1978) 432--444}.

\bibitem{HITCHIN19741}
N.~Hitchin,\,
  \href{http://dx.doi.org/https://doi.org/10.1016/0001-8708(74)90021-8}{\emph{Advances
  in Mathematics} {\bfseries 14} (1974) 1--55}.

\bibitem{Balasubramanian:1999re}
V.~Balasubramanian and P.~Kraus,\,
  \href{http://dx.doi.org/10.1007/s002200050764}{\emph{Commun. Math. Phys.}
  {\bfseries 208} (1999) 413--428}.

\bibitem{Hawking:1982dh}
S.~W. Hawking and D.~N. Page,\,
  \href{http://dx.doi.org/10.1007/BF01208266}{\emph{Commun. Math. Phys.}
  {\bfseries 87} (1983) 577}.

\bibitem{Zwiebach:1985uq}
B.~Zwiebach,\,
  \href{http://dx.doi.org/10.1016/0370-2693(85)91616-8}{\emph{Phys. Lett. B}
  {\bfseries 156} (1985) 315--317}.

\bibitem{Deser:1986xr}
S.~Deser and A.~N. Redlich,\,
  \href{http://dx.doi.org/10.1016/0370-2693(86)90177-2}{\emph{Phys. Lett. B}
  {\bfseries 176} (1986) 350}.

\bibitem{Duff:1986pq}
M.~J. Duff, B.~E.~W. Nilsson and C.~N. Pope,\,
  \href{http://dx.doi.org/10.1016/0370-2693(86)91232-3}{\emph{Phys. Lett. B}
  {\bfseries 173} (1986) 69--72}.

\bibitem{Cano:2021rey}
P.~A. Cano and A.~Ruip\'erez,\,
  \href{http://dx.doi.org/10.1103/PhysRevD.105.044022}{\emph{Phys. Rev. D}
  {\bfseries 105} (2022) 044022}.

\bibitem{Stelle:1976gc}
K.~S. Stelle,\, \href{http://dx.doi.org/10.1103/PhysRevD.16.953}{\emph{Phys.
  Rev. D} {\bfseries 16} (1977) 953--969}.

\bibitem{Starobinsky:1980te}
A.~A. Starobinsky,\,
  \href{http://dx.doi.org/10.1016/0370-2693(80)90670-X}{\emph{Phys. Lett. B}
  {\bfseries 91} (1980) 99--102}.

\bibitem{Davies:1977ze}
P.~C.~W. Davies, S.~A. Fulling, S.~M. Christensen and T.~S. Bunch,\,
  \href{http://dx.doi.org/10.1016/0003-4916(77)90167-1}{\emph{Annals Phys.}
  {\bfseries 109} (1977) 108--142}.

\bibitem{Birrell:1982ix}
N.~D. Birrell and P.~C.~W. Davies,\, , \emph{{Quantum Fields in Curved Space}}.
\newblock Cambridge Univ. Press, 2, 1984,
  \href{http://dx.doi.org/10.1017/CBO9780511622632}{10.1017/CBO9780511622632}.

\bibitem{Oliva:2011np}
J.~Oliva and S.~Ray,\,
  \href{http://dx.doi.org/10.1088/0264-9381/29/20/205008}{\emph{Class. Quant.
  Grav.} {\bfseries 29} (2012) 205008}.

\bibitem{Anabalon:2018rzq}
A.~Anabal\'on and J.~Oliva,\,
  \href{http://dx.doi.org/10.1007/JHEP04(2019)106}{\emph{JHEP} {\bfseries 04}
  (2019) 106}.

\bibitem{Anabalon:2020loe}
A.~Anabal\'on, B.~de~Wit and J.~Oliva,\,
  \href{http://dx.doi.org/10.1007/JHEP09(2020)109}{\emph{JHEP} {\bfseries 09}
  (2020) 109}.

\bibitem{Hanson:1978uv}
A.~J. Hanson and H.~Romer,\,
  \href{http://dx.doi.org/10.1016/0370-2693(78)90306-4}{\emph{Phys. Lett. B}
  {\bfseries 80} (1978) 58--60}.

\bibitem{Ghodsi:2014hua}
A.~Ghodsi, B.~Khavari and A.~Naseh,\,
  \href{http://dx.doi.org/10.1007/JHEP01(2015)137}{\emph{JHEP} {\bfseries 01}
  (2015) 137}.

\bibitem{Anastasiou:2016jix}
G.~Anastasiou and R.~Olea,\,
  \href{http://dx.doi.org/10.1103/PhysRevD.94.086008}{\emph{Phys. Rev. D}
  {\bfseries 94} (2016) 086008}.

\bibitem{Maldacena:2011mk}
J.~Maldacena,\,  \href{https://arxiv.org/abs/1105.5632}{{\ttfamily 1105.5632}}.

\bibitem{Wald:1993nt}
R.~M. Wald,\, \href{http://dx.doi.org/10.1103/PhysRevD.48.R3427}{\emph{Phys.
  Rev. D} {\bfseries 48} (1993) R3427--R3431}.

\bibitem{Iyer:1994ys}
V.~Iyer and R.~M. Wald,\,
  \href{http://dx.doi.org/10.1103/PhysRevD.50.846}{\emph{Phys. Rev. D}
  {\bfseries 50} (1994) 846--864}.

\bibitem{Padilla:2012ze}
A.~Padilla and V.~Sivanesan,\,
  \href{http://dx.doi.org/10.1007/JHEP08(2012)122}{\emph{JHEP} {\bfseries 08}
  (2012) 122}.

\end{thebibliography}\endgroup
\bibliographystyle{JHEP}

\end{document}